\documentclass[11pt]{article}
\usepackage{times,aas2pp4,astrobib,epsf}

\input{epsf}

\let\oldfootsep=\footnotesep
\def\msun { \rm {M_\odot}} 
\def\mlens { \rm {M_{lens}}} 

\def\mearth { \rm {M_\oplus}}

\def\t0{t_{\rm 0}}

\def\ie{{\it i.e. }}
\def\mlens {M} 
\def\spose#1{\hbox to 0pt{#1\hss}}
\def\simlt{\mathrel{\spose{\lower 3pt\hbox{$\mathchar"218$}}
     \raise 2.0pt\hbox{$\mathchar"13C$}}}
\def\simgt{\mathrel{\spose{\lower 3pt\hbox{$\mathchar"218$}}
     \raise 2.0pt\hbox{$\mathchar"13E$}}}

\begin{document}

\title{Simulation of a Space-Based Microlensing Survey for Terrestrial 
Extra-Solar Planets
}
\author{David P.~Bennett \& Sun Hong Rhie}      
\affil{Department of Physics, University of Notre Dame, 
                 Notre Dame, IN 46556}
\affil{email: bennett@nd.edu, srhie@nd.edu}


\vspace{-5mm}
\begin{abstract} 

We show that a space-based gravitational microlensing survey for 
terrestrial extra-solar
planets is feasible in the near future, and could provide a nearly complete
picture of the properties of planetary systems in our Galaxy.
We present simulations of such a survey
using a $1-2$m aperture space telescope
with a $\sim 2$ square degree field-of-view which is used to continuously
monitor $\sim 10^8$ Galactic bulge main sequence stars. The microlensing
techniques allows the discovery of low mass planets with high signal-to-noise,
and the space mission that we have studied are sensitive to planets 
with masses as low as that of Mars. By targeting main sequence source stars,
which can only be resolved from space, the space-based microlensing survey
is able to detect enough light from the lens stars to determine the spectral
type of one third of the lens stars with detected planets, including 
virtually all of the F, G, and K stars which comprise one quarter of 
the event sample. This enables the determination of the planetary 
masses and separations in physical units, as well as the abundance of 
planets as a function of stellar type and distance from the Galactic center.
We show that a space-based microlensing planet search program has its 
highest sensitivity to planets at orbital separations of 0.7-10 AU, 
but it will also have significant sensitivity at larger separations and will 
be able to detect free-floating planets in significant numbers. This
complements the planned terrestrial planet transit missions which are
sensitive to terrestrial planets at separations of $\leq 1\,$AU.
Such a mission should also detect $\sim 50,000$ giant planets via transits,
and it is, therefore, the only proposed planet detection method that 
is sensitive to planets at all orbital radii. 

\end{abstract}
\vspace{-5mm}

\newpage
\section{Introduction}
\label{sec-intro}

The discovery of the first extra-solar planets a few years ago 
(\citeNP{peg51,marcy-but,but-marcy}) has
spurred the growth of a new branch of observational astronomy, the study
of extra-solar planets. The success of the precision radial velocity technique
has been spectacular (\citeNP{marcy-etal,perryman,marcy-but-rev}) with the 
discovery of more than 70 extra-solar giant planets in the past seven years.
This technique is sensitive enough to detect Jupiter-mass planets in
Jupiter-like orbits, and it is anticipated that such planets will be
discovered in the next few years as the duration of the radial velocity
monitoring programs approaches Jupiter's orbital period of 12 years.
The dramatic success of these radial velocity extra-solar planet search
programs has encouraged the astronomical community to address the far more
ambitious goal of searching for Earth-like extra-solar planets
\cite{hst+beyond} because such planets seem best suited for life.
The search for Earth-like extra-solar planets has now become a major NASA
goal. It is likely that it will require the development of new
extra-solar planet search techniques since it is thought that the intrinsic
radial velocity noise of stars will limit this technique to 
planets with masses $\simgt$ a few $\times 10^{-4}$ of the host star's
mass which is 100 times greater than an Earth mass.

A number of extra-solar planet search methods have been proposed that
should be able to detect planets in the Earth mass range
\cite{perryman}. 
The Space Interferometry Mission (SIM) \cite{sim} 
will be able to detect planets of a few Earth masses around nearby 
stars via their astrometric effects on the stars they orbit. But, 
SIM requires some technical development before it will be ready to fly,
and it is not required to have the capability to detect Earth mass
planets.  The most ambitious planet search missions being considered
are the Terrestrial Planet Finder (TPF) \cite{tpf} 
and Darwin \cite{darwin} missions
which will have the ability to directly detect Earth-like planets
around nearby stars. However, these missions require a considerable amount
of technological development before they will be ready to fly. Also,
the McKee-Taylor Decadal Survey Committee \cite{decade} qualified
its endorsement of the TPF mission with the condition that the abundance
of Earth-size planets be determined prior to the start of the TPF mission.

The gravitational microlensing and transit techniques are two methods that
have sensitivity to terrestrial planets, but
are technically easier than SIM or TPF. These missions are sensitive to
planets orbiting distant stars, so they are most useful for obtaining
statistical information regarding the abundance of planetary systems.
The transit technique is employed by the COROT mission \cite{corot} which
is slated for launch by CNES in 2004, the the Eddington mission \cite{eddington}
which has been selected as an ESA F2/F3 ``reserve" mission, and
Kepler mission \cite{kepler} which is under development for NASA's
Discovery Program. However, these surveys share the property
that the transit signal due to an Earth-like planet is a photometric
variation of only $\sim 0.01\,$\%. This is only a few times above the
anticipated photometric noise, so these mission generally require the
observation of 3 or 4 transits in order to avoid false detections
due to photometric noise. Even then, 
one false detection is expected over the course of the mission
\cite{keplerweb}. The requirement for 3-4 transits limits the sensitivity of
the transit technique to planets with orbital periods of $\simlt 1$ year
due to the limited mission duration and low transit probability for
planets in longer period orbits. In contrast, the sensitivity of a
space-based microlensing planet search program extends from about $0.7\,$AU
to infinity, with significant sensitivity to free-floating planets. Thus,
the combination of a transit survey like Kepler with a space-based 
microlensing planet search will determine the abundance of terrestrial
and larger planets at all orbital radii. 

Knowledge of the general properties of planetary systems is important even
if we are primarily interested in habitable planets because the
issue of planetary habitability is a complex and poorly defined one. 
The Earth's habitability is a consequence of a complex interplay of physical 
processes (Lunine 1999) that are not likely to be replicated in exactly 
the same way on other worlds. While the fundamental requirement is assumed to
be stable liquid water over geologic time, many diverse factors come 
into play in establishing habitable ecosystems (Des Marais et al. 2001). 
More importantly, we do not know what the outcome of a different 
combination or timing of such processes would be in terms of 
habitability (Chyba et al. 2000). A non-exhaustive list of the 
potential requirements for habitability include the presence of 
giant planets in ~5-10 AU orbits (Lunine 2001), the presence of 
a large moon to stabilize the planetary spin axis (Ward 1982), 
main sequence stellar type of F, G, or K (Ward and Brownlee 2000; 
Kasting 1997). Also, the traditional notion that a narrow range of 
semi-major axes are consistent with the presence of liquid water 
(Kasting, Whitmire \& Reynolds 1993) is challenged by the evidence 
for liquid water on the early Mars (Carr 1996). The length and 
incompleteness of this shopping  list demands survey missions 
be initiated soon to map out the geometries of extra-solar 
planetary systems prior to much more expensive missions whose 
intent is to spectroscopically examine extra-solar terrestrial 
planets. With its high sensitivity to low-mass planets at a wide 
range of separations, a space-based gravitational microlensing
survey would be the ideal mission for a comprehensive 
survey of the properties of planetary systems.

\subsection{The Gravitational Microlensing Technique}
\label{subsec-microlens}

The gravitational microlensing technique 
(\citeNP{mao-pac,gould-loeb,dps2000-micro}), has the unique property
that the strength of the planet's photometric microlensing signal is nearly
independent of the planetary mass. Instead of a weaker signal, the 
microlensing signals of low-mass planets have a shorter duration and 
a lower detection probability than those of high-mass planets. (This
argument breaks down for planetary masses below $0.1\mearth$ because
such planets lens only a fraction of the main sequence source star disks.)
This means that a microlensing survey with frequent observations of
a very large number of stars will be able to detect terrestrial planets
at high signal-to-noise 
(\citeNP{tytler-exnps,bennett-rhie,wam,bennett-rhie-2k}).
The microlensing technique employs stars in the Galactic bulge which
act as sources of light rays which are bent by the gravitational fields
of stars in the foreground: on the near side of the Galactic bulge, or in
the disk. Planets which may orbit these ``lens" stars can be detected
when the light rays from one of the lensed images pass close to a planet
orbiting the lens star. The gravitational field of the planet distorts
this lensed image causing a significant variation of the gravitational
microlensing light curve from the standard single lens light curve.
This planetary deviation is typically of order $\sim 10$\%, and it
has a duration of a few hours to a day compared to the typical 1-2 
month duration for lensing events due to stars.

The main challenge for a microlensing planet search project is that
microlensing events are rare. Only about $3\times 10^{-6}$ of Galactic bulge 
stars are microlensed at any given time 
\cite{ogle-tau,macho-bulge45,macho-bulge-diff}, and only $\sim 2$\% of 
earth-mass planets orbiting these stars will be in the right position to 
be detected \cite{bennett-rhie}. 
The sensitivity limit of the gravitational microlensing technique is set by the
finite angular size of the source stars because a very low mass planet
will only deflect the light rays from a fraction of the source star's disk. 
This can wash out the photometric signal of the planet.
For main sequence source stars in the Galactic bulge, the
sensitivity limit is about $0.1 \mearth$, but for giant source stars, it is
$> 1 \mearth$. Thus, a gravitational microlensing search for terrestrial
planets must use main sequence source stars. However, the density of 
bright main sequence stars in the central Galactic bulge is several
stars per square arc second, so angular resolution of $\ll 1\,$arc sec
is necessary to resolve these stars. 

In order to accurately characterize the
parameters of the planets discovered via microlensing
\cite{gaudi-gould,gaudi}, we must have
photometry of $\sim 1$\% accuracy sampled several times per hour over
a period of several days (\ie\ a factor of a few longer than the
planetary light curve deviation). The microlensing event light curves
must also be sampled continuously for periods of more than 24 hours,
in order to unambiguously characterize the planetary signals in 
microlensing light curves.  This allows
both the full planetary deviation as well as the periods before and after it
to be observed. 

\subsection{Ground-based Microlensing Planet Searches}
\label{subsec-ground}

The earliest discussions of detecting terrestrial planets
via gravitational microlensing 
generally considered it to be a technique for ground-based observations
(\citeNP{tytler-exnps,bennett-rhie,wam}). However, these early estimates
proved to be overly optimistic in a number of respects. \citeN{peale}
performed a simulation of what sort of planets could be detected by
a global network of $\sim 2\,$m telescopes as suggested by \cite{tytler-exnps}
showed that a substantial number of possible planet detections would
be missed due to poor weather and geographic limitations on the locations
of ground based telescopes, but his results are 
over-optimistic for several different reasons. For example,
no account was taken of variations in atmospheric
seeing or of the poorer average seeing from the non-Chilean observing sites.
\citeN{sackett97} sought to avoid the problems of the poorer observing sites
by proposing a search employing only a single, excellent observing site,
Paranal under the assumption that the planetary signals of terrestrial
planets would be brief enough that some could be fully characterized by
observations spanning $\simlt 8\,$hours.

\begin{figure}
\plotone{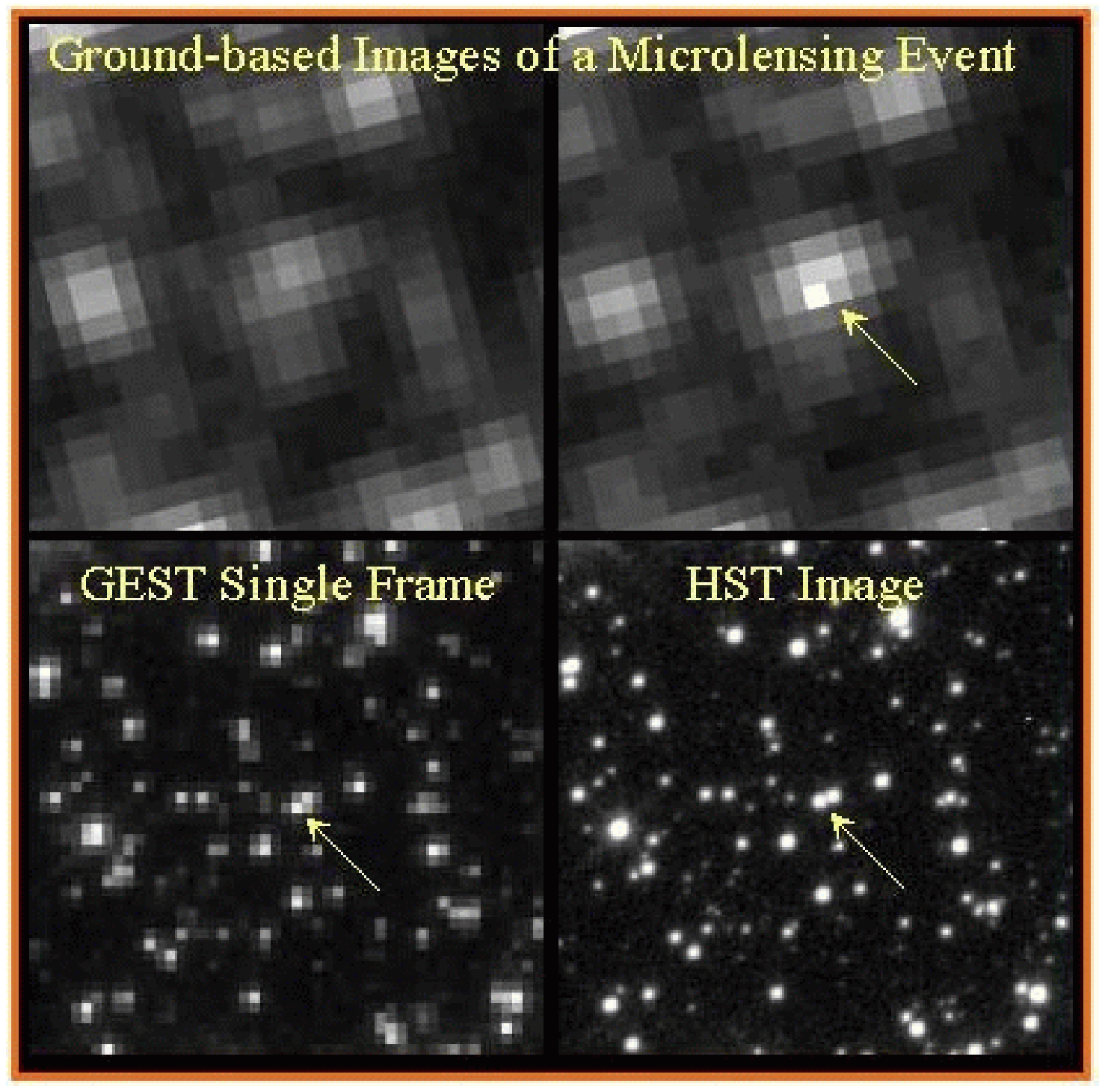}
\caption{The difference between ground and space-based data for microlensing
of a bulge main sequence star is illustrated with images of microlensing
event MACHO-96-BLG-5. The two top panels are 50 min. R-band exposures 
with the CTIO 0.9m telescope taken in 1" seeing at different microlensing 
magnifications, and the two images on the bottom have been constructed 
from HST frames. The bottom left image represents a 10 minute exposure 
with GEST's angular resolution and pixel size, and the image on the 
right is an HST image. The lensing magnification factors are 
A = 4 and 10 for the ground based images and 1.07 for the
space based image. The source star, a Galactic bulge G-dwarf is 
indicated by the yellow arrows.
\label{fig-image}}
\end{figure}

All of these papers considered the monitoring Galactic bulge turn-off
stars. Ground-based color magnitude diagrams of the dense Galactic bulge
fields observed by the microlensing surveys seemed to show that there
were very large numbers of these stars. 
Turn-off stars are stars that have recently exhausted the
Hydrogen fuel in their cores, and are just beginning the Hydrogen shell
burning phase. They are 1-2 magnitudes brighter than the stars at the
top of the main sequence, but have similar colors. Their radii are small
enough to allow the detection of Earth-mass planets via microlensing.
However, this is a relatively brief phase of stellar evolution, and so
their apparent abundance in Galactic bulge seemed odd. In fact, this
abundance was not confirmed with HST data \cite{holtzman}. The apparent
abundance of these ``turn-off" stars is an artifact the stellar crowding
in these central Galactic bulge fields; the density of main sequence stars
is too high for them to be individually resolved, and several main sequence
stars blended together are typically identified as a single ``turn-off"
star. This is illustrated in Fig.~\ref{fig-image} which compares two
ground based images of microlensing event MACHO-96-BLG-5 to an HST image
and a simulated image from a proposed space-based
microlensing planet search telescope. 
Clearly, density of bright main sequence stars
(like the one indicated) is too high for these stars be individually
resolved from the ground.
This stellar blending phenomena has been widely discussed
in the gravitational microlensing literature 
(\citeNP{distef-esin,woz-pac,han}), and there is now strong evidence 
that virtually all of 
the microlensing events involving apparent bulge ``turn-off" source stars 
are, in fact, blended microlensing events with main sequence source stars
(\citeNP{macho-binaries,macho-bulge-diff}). This blending of source stars
makes microlensing planet searches much more difficult because the planetary
signal will be confined to the flux from only one of the blended
stars, but all of the blended stars will contribute to the photometric noise.

One can hope to compensate for the increased photometric noise caused by
blended by moving to relatively large wide field-of-view ground-based 
telescopes, such as the $4\,$m VISTA telescope or the $\sim 8\,$m LSST 
(the Large Synoptic Survey Telescope - recommended by the McKee-Taylor 
Committee \cite{decade}). A detailed study of such potential observing
programs reveals that a survey from an excellent observing site such as
Paranal is about two orders of magnitude less sensitive than a space-based
microlensing planet search program \cite{bennett-rhie-vista}. A critical
problem for such a ground-based program is that the typical duration of
a microlensing light curve deviation due to an Earth-mass planet is
nearly 24 hours. This is about an order of magnitude longer than 
the Einstein radius crossing time, which was used as the characteristic
planetary event duration in a previous study which advocated a microlensing
planet survey from a single site \cite{sackett97}. With a realistic
distribution of event durations, however, we find that only a very small
subset of Earth-mass planetary microlensing signals can be detected and
characterized. The detectable Earth-mass planet events from such a survey
also suffer from several undesirable selection effects. There is a much
higher fraction of high magnification events with planetary separations
very close to the Einstein radius, and these events provide essentially
no information on the abundance of planets as a function of orbital separation.
Finally, few of the events which are detected with these ground based
surveys allow the detection of the lens star, so the planetary abundance as
a function of spectral type also cannot be measured from the ground.

\subsection{Microlensing Planet Search Space Mission Requirements}
\label{subsec-gest}

The primary
intent of this paper is to investigate low cost space missions which 
employ the gravitational microlensing technique to
detect terrestrial planets orbiting other stars.
The basic requirements for such a mission are
that $\sim 10^8$ Galactic bulge main sequence stars must
be observed almost continuously at intervals of 20 minutes, or less for 
periods of at least several months. Photometric accuracy of $\sim 1\,$\% or
better is needed, and this implies that the angular resolution of the images
must be $< 0.4$" in order to resolve main sequence stars in the crowded
central Galactic bulge fields. The required frequent photometric measurements
of such a large number of stars requires relatively high data rate of
$\simgt 10\,$Mbits/sec depending upon the data compression scheme
that is used.

While the wide-field imaging capabilities required for such a mission 
substantially exceed the capabilities of existing space telescopes, it
it can be undertaken at a relatively modest cost, within the limits of
NASA's MIDEX or Discovery programs. There may also be
an opportunity to combine a microlensing planet search mission with another
major science program, such as 
a deep, wide field, weak gravitational lensing survey or
a high-redshift Supernova search similar to the proposed SuperNova
Acceleration Probe\footnote[1]{See {\tt http://snap.lbl.gov} for information
regarding the proposed SNAP mission} (SNAP). 
It might also be possible to combine
a microlensing planet search mission with an asteroseismology program such as
the Eddington mission\footnote[2]{See 
{\tt http://astro.esa.int/SA-general/Projects/Eddington/} 
for information on  ESA's Eddington mission.}
mission if such a mission were designed with good in-focus optics. 

Two proposals for microlensing planet search missions have been submitted to
NASA in 2001. The Galactic Exoplanet Survey Telescope 
(GEST\footnote[3]{
More information on the Galactic Exoplanet Survey Telescope is available at
{\tt http://bustard.phys.nd.edu/GEST/}})
was submitted to NASA's MIDEX Program, and the Survey for Terrestrial
ExoPlanets (STEP) was submitted to NASA's Extra Solar Planets:
Advanced Concepts program. We will use the GEST MIDEX proposal as the baseline
for our discussions of planet detection sensitivity, but we will also 
investigate the variation of the planet detection sensitivity on the
parameters of the mission.

The terrestrial planetary signals in gravitational microlensing light curves
that these missions would study show significant variations on time scales
ranging from 20-30 minutes to about a day. Therefore, it is important that
a microlensing planet search telescope be in an orbit which allows 
continuous viewing of the Galactic bulge. The orbit proposed for GEST is a
polar orbit with an altitude of $\sim 1200\,$km oriented to keep
the Galactic bulge in the continuous viewing zone, while the STEP
mission would employ a nearly circular geosynchronous orbit 
inclined by $28.7^\circ$ (the latitude of Cape Canaveral)
from the equator and by $\sim 50^\circ$ with respect to the ecliptic plane.
Even higher Earth orbits, such as the 14-day ``Prometheus"
orbit proposed for SNAP, would also be acceptable, but Earth-trailing 
orbit might make it difficult to achieve the required data rate.

\begin{figure}
\plotone{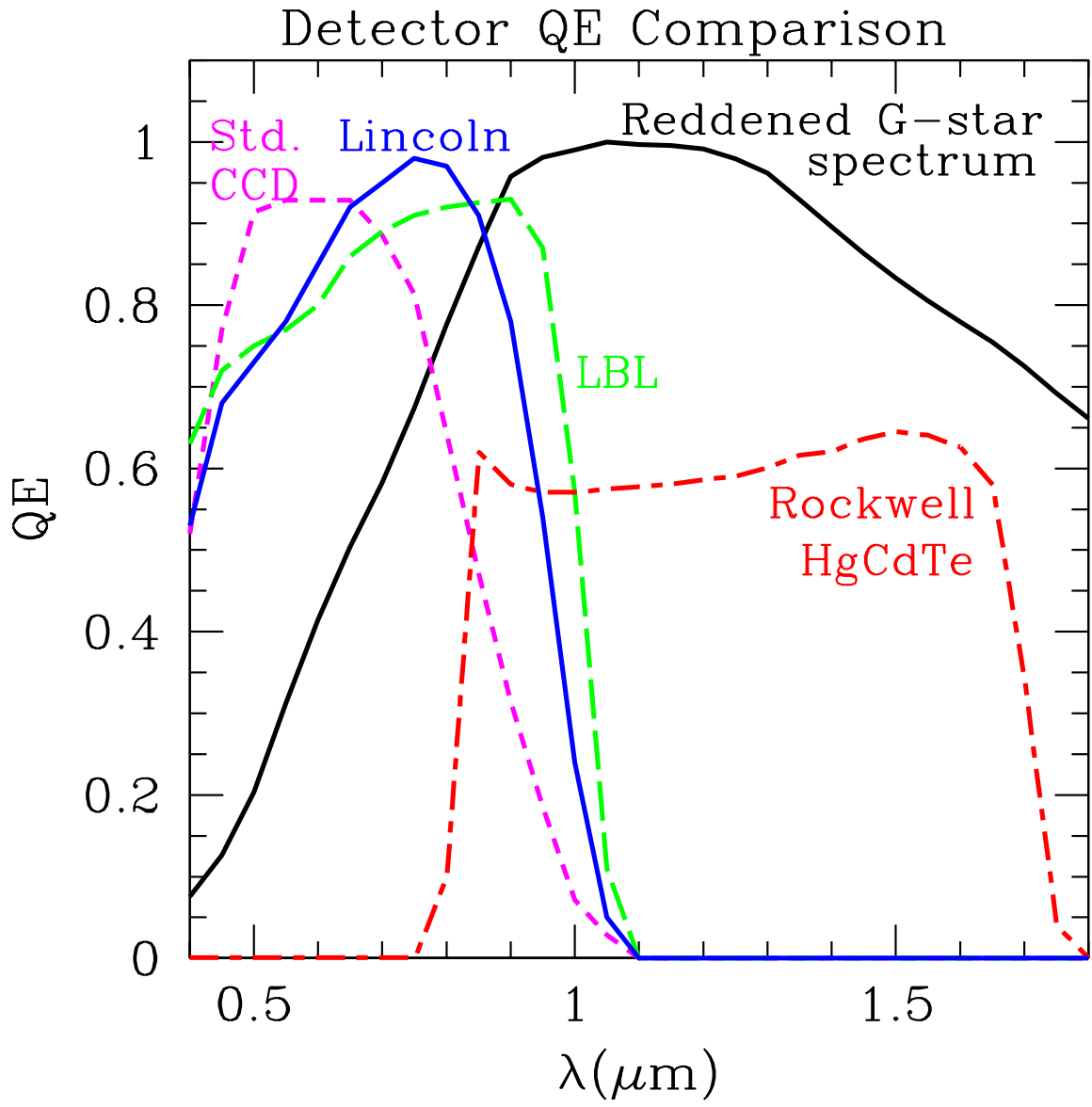}
\caption{The spectrum of a typical reddened, Galactic bulge
source star is compared to the
quantum efficiency curves for detectors that might be used for a 
microlensing planet search program. The Std. CCD curve represents a Marconi
CCD, which has a QE curve similar to the CCDs that are currently being produced
by Fairchild and SITe as well as Marconi. The Lincoln Labs and LBL CCDs
use high resistivity Silicon for enhanced sensitivity in the near IR, and
the Rockwell device is one designed for HST's upcoming Wide Field Camera 3.
\label{fig-qe}}
\end{figure}

The GEST and STEP designs call for $1.0$m and $1.5$m aperture telescopes
each with a 2.2 square degree field of view
and a three mirror anastigmat design. The field of view is elliptical
with an axis ratio of about 2:1, and the GEST proposal would use
an array of 32 $3072\times 6144$ pixel Lincoln Labs high resistivity CCDs
for enhanced sensitivity in the near IR. The STEP proposal would use a 
combination of these same Lincoln Labs CCDs and Rockwell HgCdTe IR detector
arrays. These IR detectors would be similar to a design intended for
the Hubble Space Telescope's Wide Field Camera 3 with a long wavelength
cut-off of $\sim 1.7\,\mu$ to allow radiative cooling from a high Earth
orbit. The quantum efficiency of these detectors is
displayed in Fig.~\ref{fig-qe} along with the reddened spectrum of
a typical bulge source star. The standard CCD curve is typical of 
most broadband astronomical CCD detectors (such as those manufactured by 
Marconi, SITe, or Fairchild). Both the Lincoln Labs and
LBL devices use high resistivity Silicon to enhance sensitivity in the
near IR, and the LBL devices have higher sensitivity near $\lambda = 1.0\mu$
because they are more opaque at this wavelength due to their $300\mu$ 
thickness vs.~$50\mu$ for the Lincoln devices. 

The overall sensitivity of the detectors is given by the integral of 
the product of the source spectrum, the QE curve, and an additional
function describing the throughput of the rest of the optical
system. If we assume that the optical system has no other significant
wavelength dependence besides the detectors, then we find sensitivity
improvements of 44\%, 62\%, and 150\% for the Lincoln, LBL, and Rockwell
detectors, respectively when compared to the sensitivity of the
standard astronomical CCD. At the time of this writing, only the
standard and Lincoln CCDs can be produced in the large quantities needed
for a microlensing planet search mission, but this situation may change
in the near future.

It is anticipated that GEST and STEP will 
take $\sim 100$ second exposures at
2 minute intervals which would be co-added into ten minutes exposures.
Assuming digitization at 14 bits/pixel, this gives a total data rate of
$14\,$Mbits/sec if no data compression is employed.

In order to make use of the high angular resolution available from space,
it is necessary for the space-based telescope to have high pointing
stability. We require that the pointing be stable to 10\% or better of the
assumed 0.2 arc second CCD pixel size. This should be achievable with
three-axis stabilized, ultra-low jitter spacecraft, such as Lockheed's
LM-900, as long as a fine-guidance signal is provided from guide CCDs in
the focal science focal plane which would be read out a few times a second.
Another option, which could correct higher frequency pointing jitter, would
be to use a fast-guiding secondary mirror.
The only pointing variation needed during the $\sim 8$ month
Galactic bulge season would be a sub-pixel scale dither pattern needed
to ensure that the photometric accuracy remains very close to the
photon noise limit (\citeNP{lauer,gilliland}).

In this paper, we present the results of a detailed simulation of 
a space-based microlensing planet search mission. 
In section \ref{sec-details}, we explain the assumptions and
the details of our simulation and argue that our assumptions are conservative.
In section \ref{sec-results}, we present the details of our results
including example light curves, the predicted planet detection 
sensitivity to bound and free-floating planets, and the prospects
for direct observations of the lens stars. There is also a brief discussion
of the $\sim 50,000$ planets that such a mission
is likely to detect via transits.
Finally, in section \ref{sec-conclude}, we summarize the scientific
results to be expected from a space-based microlensing planet search mission.

\section{Mission Simulation Details}
\label{sec-details}

In order to simulate a space-based microlensing planet search mission, 
we must make assumptions regarding
the source stars, the lens star systems and the telescope. Our
distribution of source stars is based upon the Galactic bulge luminosity
function of \citeN{holtzman}. Following the GEST proposal, we select
a field at Galactic
coordinates $l\approx 1.2^\circ$, $b\approx -2.4^\circ$, which is closer to
the Galactic Center than the Baade's window field observed by 
Holtzman et~al. This implies that both the star density and the reddening 
will be higher, and we split the field into two pieces for the purposes
or our simulations in order to account for the gradient of the star density
with Galactic latitude. The two half-fields have central Galactic
latitudes of $b = -2.0^\circ$ and $-2.8^\circ$, and we have assigned them
star densities of 2.06 and 1.55 times the \citeN{holtzman} star density
measured at $b = -3.9^\circ$ based upon number counts of ``red clump"
stars in the MACHO fields \cite{piotr}. The I-band extinctions for these two
half fields are assumed to be $A_I = 1.6$ for the inner half field and
$A_I = 1.5$ for the outer half field. These reddening values can be
obtained from the \citeN{dust-map} dust map with a correction for
stellar emission as advocated by \citeN{stanek} or by assuming that the
excess IR emission is proportional to the ``red clump" star number counts.

Another very crucial physical input for our simulation is the
microlensing probability (or optical depth, $\tau$) towards the Galactic bulge.
Measured values are $\tau = 3.3\pm 1.2 \times 10^{-6}$ at $l = 0.9^\circ$,
$b = -3.8^\circ$ \cite{ogle-tau}, 
$\tau = 3.9 {+ 1.8 \atop - 1.2} \times 10^{-6}$ at 
$l= 2.55^\circ$ and $b = -3.64^\circ$ \cite{macho-bulge45}, and
$\tau = 3.23 {+ 0.52 \atop 0.50} \times 10^{-6}$ 
at $l= 2.68^\circ$ and $b = -3.35^\circ$ \cite{macho-bulge-diff}. 
We have used this latest measurement because it is based
upon the largest sample, and it is closest to the theoretical estimates.
Theoretical determinations of the scaling of the microlensing probability
with position \cite{bissantz,peale-tau} indicate that the microlensing 
probability at GEST's outer half field ($l = 1.2^\circ$, $b = -2.8^\circ$) 
should be 1.2-1.3 times larger than at $l= 2.68^\circ$, $b = -3.35^\circ$, 
while the increase at the inner half field ($l = 1.2^\circ$, $b = -2.0^\circ$)
should be a factor 1.4-1.8. For the purposes of this simulation, we
have selected a conservative choice for the microlensing probability,
$\tau = 2.43 \times 10^{-6}$ at $l= 2.68^\circ$ and $b = -3.35^\circ$
which we then scale to $\tau = 2.9\times 10^{-6}$ at $l = 1.2^\circ$,
$b = -2.8^\circ$, and $\tau = 3.9\times 10^{-6}$ at $l = 1.2^\circ$, 
$b = -2.0^\circ$. This is the $1.6\sigma$ lower limit on the value of $\tau$
extrapolated to our selected field.

The mass function of the lens stars is assumed to follow the a
conventional power law form, $f(m) \propto m^{-\alpha}$ where
$f(m) dm$ is the number of stars in the mass interval $m$ to $m + dm$.
We use a mass function similar to those advocated by
\citeN{zoccali} and \citeN{kroupa} which imply
different values of $\alpha$ in different mass intervals:
$\alpha = 2.3$ for $m > 0.8\msun$, $\alpha = 1.33$ 
for $0.15\msun < m < 0.8\msun$, and $\alpha = 0.3$, for 
$0.05\msun < m < 0.15\msun$. The mass function is truncated at
$0.05\msun$ in order to keep the distribution of microlensing event
timescales consistent
with the observations of \citeN{macho-bulge-diff}. Stellar remnants are also 
included with white dwarfs contributing 13\% of the lens stars, while neutron
stars and black holes contribute $< 1$\% and $< 0.1$\% of the lens
stars, respectively. 

With these parameters for the properties of the inner Galaxy, we precede
to run our simulations as follows:
\begin{enumerate}
 \item We create an artificial image with stars
       $0 \leq M_I \leq 9$ at random locations in an artificial image
       using a ``pseudo-gaussian" profile (as in DOPHOT \cite{dophot})
       with a FWHM of 0.24\rlap." Brighter stars are not included, but we 
       assume that 5\% of the 2.1 square degree field of view is lost
       due to bright, saturated stars or CCD defects.
 \item A stellar lensing event is selected for each star in the frame
       with lens parameters selected at random assuming the mass
       function described above and a density and velocity distribution
       from a standard model of the Galaxy \cite{han-gould}.
       All stellar lensing events are assumed to have an impact parameter of
       $\leq 3$ Einstein radii, and the source stars are assumed to
       reside at $0.5\,$kpc behind the Galactic Center which is at
       $R_0 = 8\,$kpc.
 \item The orientation of each ``exo-ecliptic" plane is selected at random,
       and then planet locations are selected by assigning each planet
       a random orbital phase within this plane. The planets are assumed to
       follow circular orbits with radii between
       0.25 and $30\,$AU and mass fractions ranging from 
       $\epsilon = 3\times 10^{-7}$ to $\epsilon = 10^{-3}$.
 \item Planetary lensing light curves are constructed assuming measurements
       every ten minutes. Finite source effects are incorporated  assuming
       a mass radius relationship taken from \citeN{bertelli}.
 \item The CCD camera is assumed to detect
       13 photons per second from an $I = 22$ star. This can be achieved 
       with a 1.5m telescope with standard CCDs employing a 650-900nm passband,
       or with a 1.0m telescope with high-resistivity Lincoln labs or 
       LBL CCDs with a 500-1000nm passband.
 \item Light curve error bars are generated under the assumption that
       the photometric accuracy is limited by photon statistics for noise 
       levels down to 0.3\%. This level of accuracy has been demonstrated
       with highly undersampled HST images of very crowded star fields
       (\citeNP{lauer,gilliland}). The key to this photometric accuracy is to
       recover the diffraction limited resolution with a sub-pixel scale
       dither pattern. The undersampling of these HST images
       is similar to the level of undersampling for the proposed GEST
       mission, as shown in Fig.~\ref{fig-image}.
       In addition to the source star, the
       lens star and nearby stars with images that are blended
       with the source star are assumed to contribute to the photon noise.
 \item A single lens, point source light curve is fit to each event,
       and planet detections are signaled by an excess fit $\chi^2$.
       We measure the planetary signal with the $\Delta\chi^2$ which is
       the difference between the $\chi^2$ for the single lens fit and
       the correct planetary lensing fit.
       Our detection threshold is $\Delta\chi^2 \geq 160$ which is the
       equivalent of a $12.5\sigma$ detection.
\end{enumerate}
One potential drawback with our method for identifying planet detections 
is that planet detections may be incorrectly indicated for events with
very high magnification because the effects of the finite angular size
of the source star may be seen. These high magnification events also
have higher sensitivity to planets than lower magnification events
\cite{griest-saf}
because the source star must necessarily pass close to the ``stellar"
caustic curve which will be distorted due to the presence of planets.
However, the determination of the planetary mass fraction ($\epsilon$)
and separation can be difficult for events detected due to the
stellar caustic \cite{dominik}. Thus, it is not yet clear how useful
such detections will be, although they do present enhanced sensitivity to
multiple planets \cite{naber}. Because of this uncertainty, we have
excluded planets detected in events with maximum magnifications $> 200$.
 
\section{Expected Results}
\label{sec-results}

\subsection{Planetary Parameters from Microlensing}
\label{subsec-planpar}

The diversity
of microlensing planetary light curves has been studied quite extensively
(\citeNP{mao-pac,gould-loeb,bolatto-falco,bennett-rhie,wam,gaudi-gould,gaudi}),
and these studies have shown that it is possible to measure both the
planetary mass fraction, $\epsilon$, and the planet-star separation 
from the light curve shape. The duration of the planetary light curve deviation
gives $\epsilon$. The overall magnification of the light curve
at the time of the planetary deviation and the basic shape of the planetary
deviation give the separation. However, the transverse separation, $a$,
is only determined in units of the Einstein ring radius,
\begin{equation}
  R_E = 2.85\,{\rm AU}\,\sqrt{{\mlens\over \msun} {D \over 1\,{\rm kpc}}} \ ,
\end{equation}
which is just the radius of ring image for a single lens of mass $\mlens$
that is perfectly aligned with the source star. $D = D_l(D_s-D_l)/D_l$, where
$D_l$ and $D_s$ are the
distances to the lens and source stars, respectively. 

For a source star in the Galactic bulge,
$R_E$ is typically $\sim 2\,$AU, and it ranges from 1-4 AU, so a measurement
of $a/R_E$ will yield an estimate of $a$ that is good to a factor of 2.
For most of the terrestrial planet detections, however, we can do somewhat
better than this because we can also measure the time for the lens
center-of-mass to cross the source star radius, $t_s$. This parameter
is measurable for events in which the source comes very close to or
crosses one of the lens caustics. This occurs for a large fraction of the
terrestrial planet events, but there are many of the giant planet lensing
events that are detectable without a close approach to a caustic.
Precise values of $a$ and $\mlens$ can be obtained
for events in which the lens can be detected either via multi-color
photometry, spectroscopy, or
proper motion, as the lens separates from the source in the years
after the event. This should be possible for about one third of the
events including virtually all of the F, G and K star lenses.
(See subsection \ref{subsec-sstar} for a more detailed discussion of 
source star identification.)

\subsection{Event Light Curves}
\label{subsec-lc}

\begin{figure}
\plottwo{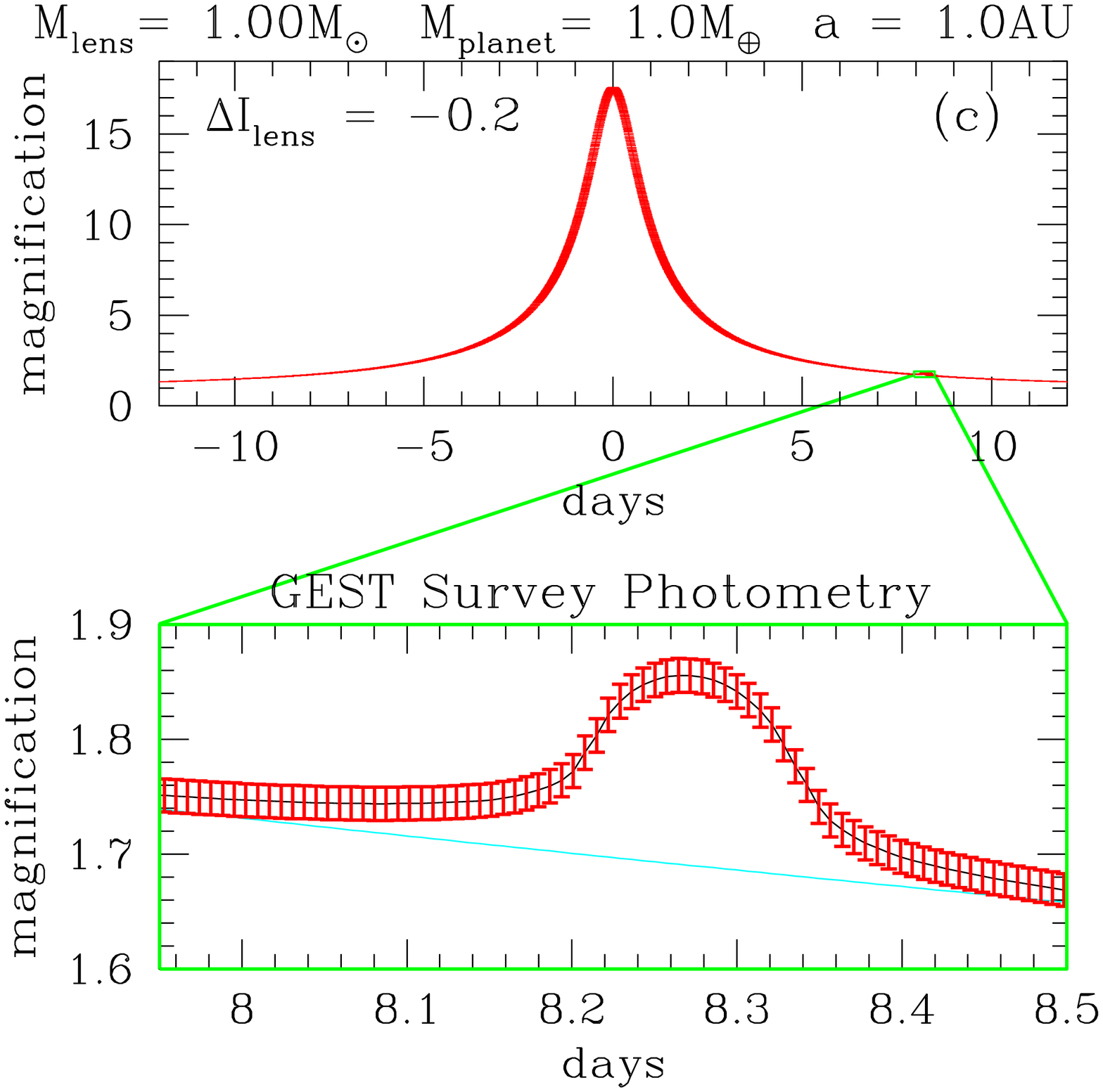}{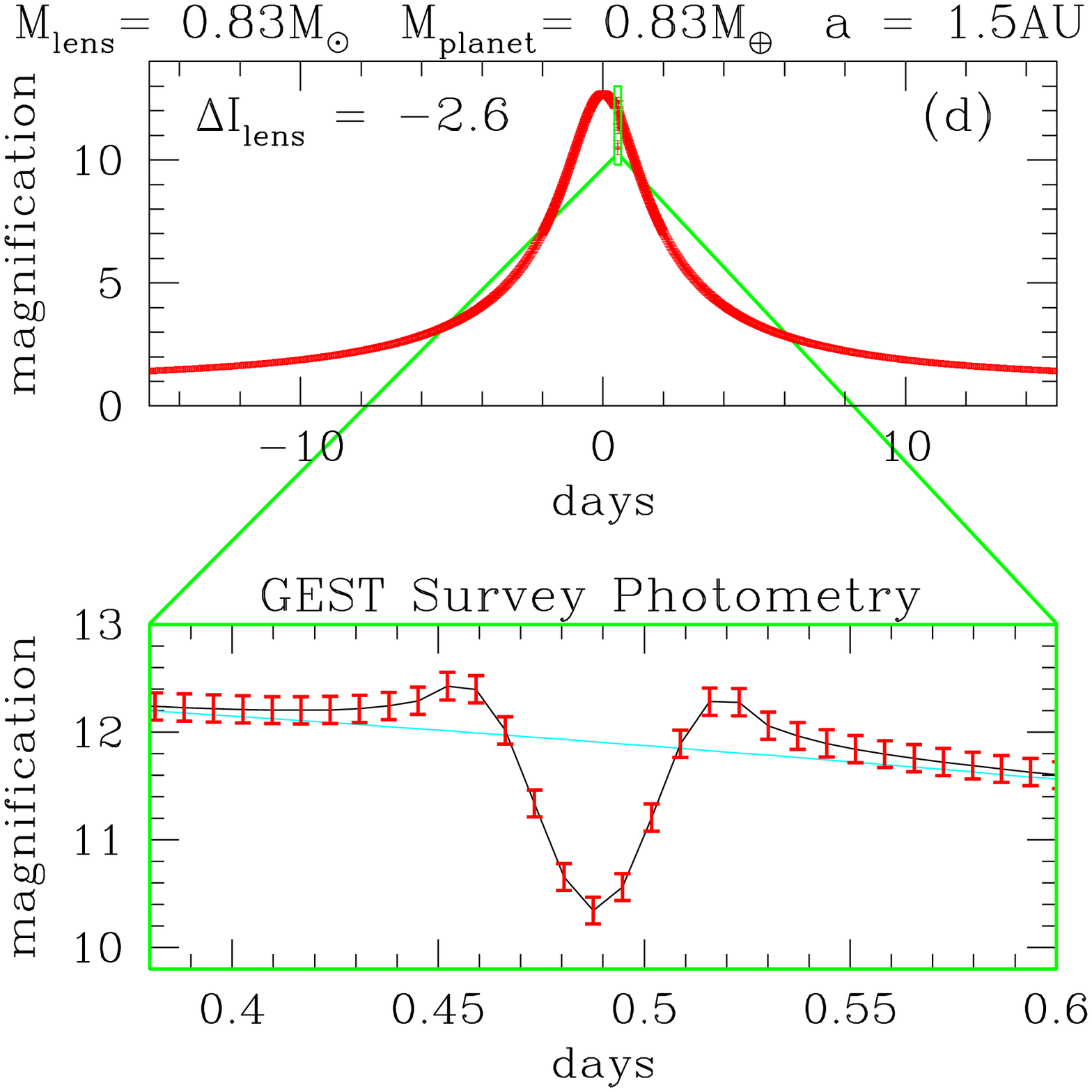}
\caption{
Example light curves are shown from a simulation of the GEST mission. In each
case, the top panel shows the full light curve, and the planetary 
deviation regions are blown up and shown in the lower panels. All of the 
example light curves have the Earth:Sun mass ration of 
$\epsilon = 3 \times 10^{-6}$.
Figs. (a) and (b) span the range of planetary 
detection significance from $\Delta\chi^2 = 60,000$ (a) to 
$\Delta\chi^2 = 180$ (b) which is close to our cut. Figures (c) and
(d) show more typical light curves with 
$\Delta\chi^2 = 3000$ and $\Delta\chi^2 = 500$, respectively. 
The planets detected in (b) and (c) have orbital radii of $1\,$AU while the
events shown in (a) and (d) have orbital radii of 5 and $1.5\,$AU, 
respectively.  $\Delta {\rm I}_{\rm lens}$
is the difference between lens and source I magnitude.
\label{fig-lc1}}
\end{figure}

\begin{figure}
\plottwo{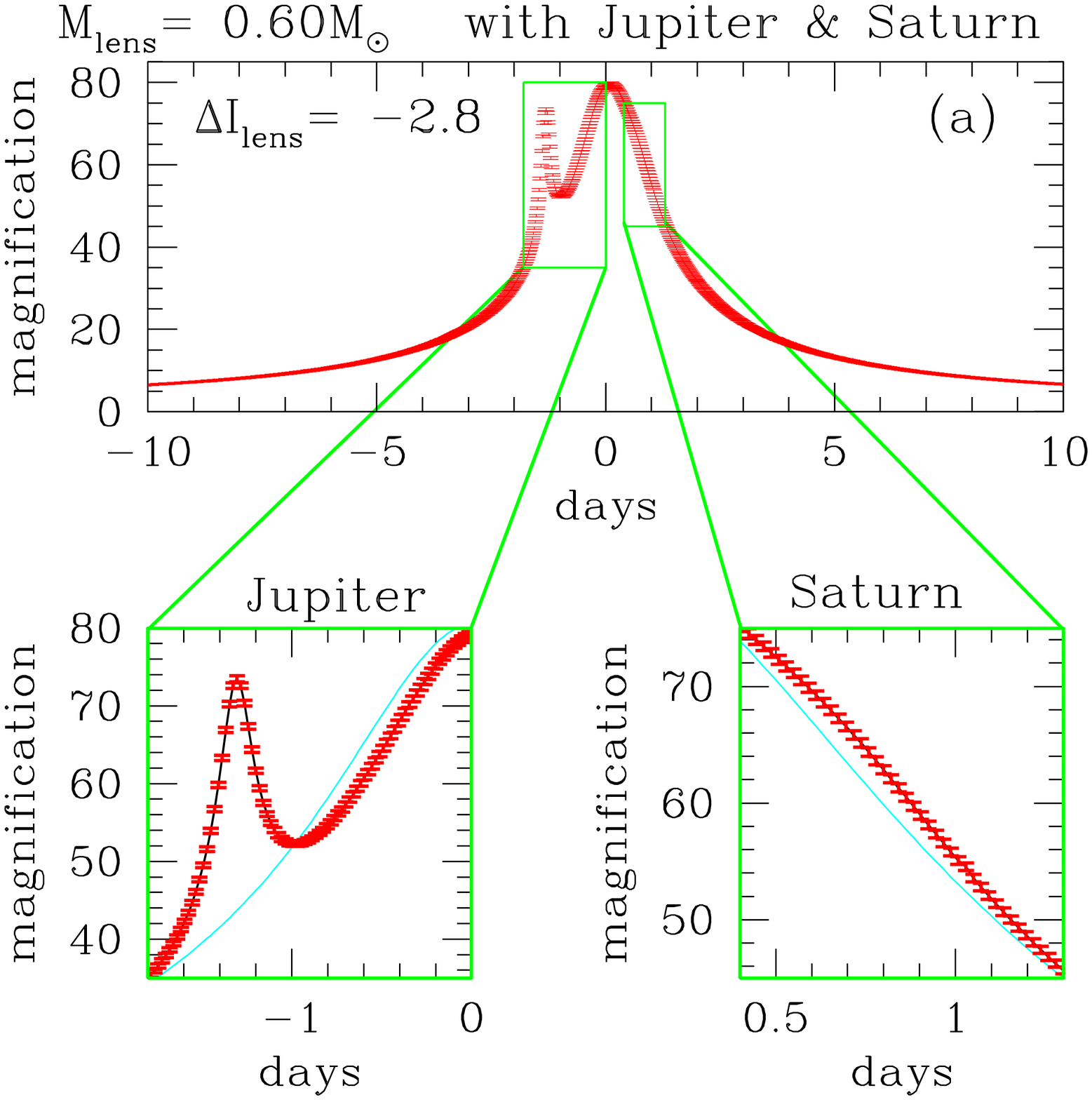}{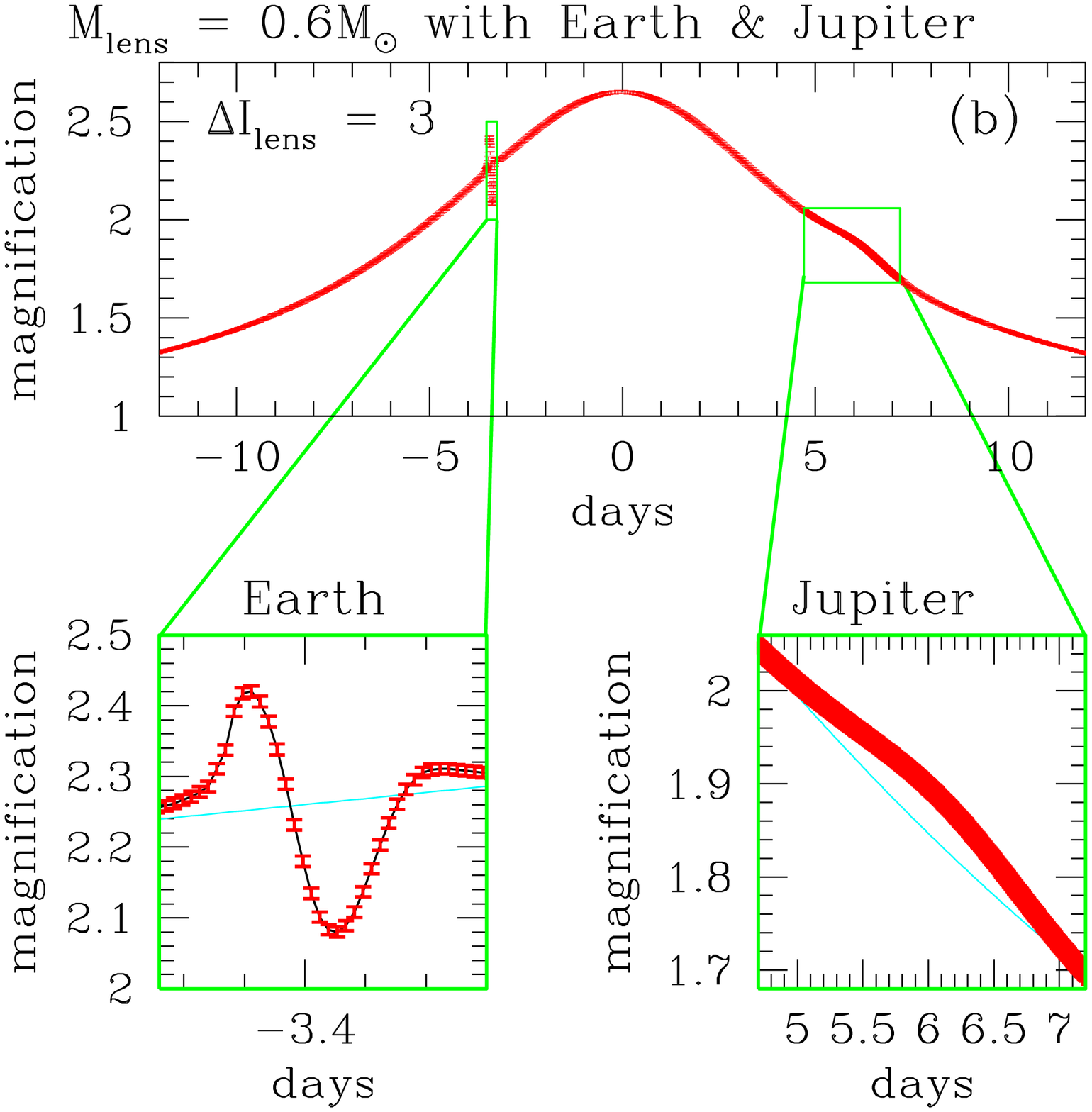}
\caption{
Example multiple planet light curves from our simulation of planetary systems
with the same planetary mass ratios and separations as in our solar system.
(a) is an example of a Jupiter/Saturn detections and 
(b) is an example of the detection of Earth and a Jupiter.
\label{fig-lc2}}
\end{figure}

Examples of the planetary light curves from our GEST mission
simulation are shown
in Figures \ref{fig-lc1}-\ref{fig-lc_moon}. The data are shown with the
error bars determined as described above, and the light curves are
presented with the sampling interval of 10 minutes that was used for
the event detection calculations. While the error bars are meant to 
indicate the $1\,\sigma$ uncertainties, we have not added this noise to
the data points shown in Figures \ref{fig-lc1} and \ref{fig-lc2} because
of the high density of data points in these figures.
These light curves are meant to illustrate
the range of planetary light curves that a space-based microlensing
survey should detect. They also
represent the range of signal-to-noise of the terrestrial
planet detections in our GEST simulations.  Figure \ref{fig-lc1}(a)
represents one of the highest signal-to-noise planet detections
with the Earth:Sun mass ratio of $\epsilon = 3\times 10^{-6}$, and 
Figure \ref{fig-lc1}(b) is an event which barely passes our
event detection cut of $\Delta\chi^2 \geq 160$. The other events have
more typical signal-to-noise.

We've assumed that the photometric accuracy of a space-based microlensing
survey will be dominated by photon
statistics and that systematic errors will not become dominant until
the statistical errors reach $< 0.3$\% (in 5 co-added $100\,$sec exposures).
However, Figures \ref{fig-lc1} and \ref{fig-lc2} illustrate that most of 
the planet detections are made with lower precision photometry, with 
photometric errors of $\sim 1$\% dominated by photon statistics.

These events serve to illustrate why ground based microlensing searches
are not effective for the detection of terrestrial planets
(\citeNP{bennett-rhie-2k,dps2000-GEST}). 
The necessity of using main sequence target stars
for a microlensing program to find terrestrial planets means that the
accuracy of photometry is compromised by the blending of the source star
images as demonstrated in Fig.~\ref{fig-image}.
This is true even if the planet search program is limit to the best
ground based observing sites such as Paranal \cite{sackett}. This
blending with neighboring stars less than an arc second away
substantially reduces the photometric
signal-to-noise and would make the events shown in Figs.~\ref{fig-lc1}(b)-(d)
undetectable. The event shown in Fig.~\ref{fig-lc1}(a) 
would have a large enough
signal to be detectable from a ground-based program, but since the planetary
deviation lasts for more than 24 hours, it would be poorly sampled from 
a single site. Follow-up observations from sites at other longitudes
would be of little help because the poorer seeing at these sites would
make the photometry too noisy to be very useful in characterizing the
properties of the detected planet.

\subsubsection{Light Curves for Multiple Planets and Moons}
\label{subsec-lcmult}

Figure \ref{fig-lc2} shows events in which multiple planets are detected.
Most multiple planet events have 
light curves that are very similar to single planet events except 
that that are two different planetary deviation regions. 
We've run simulations of ``solar-type" planetary systems in which every
stellar lens is assumed to have planets with the same mass fractions as
the planets in the solar system and with the same separations.
Most of the multiple planet detections in our simulations are similar to
Figure \ref{fig-lc2}(a) in which both the ``Jupiter" and ``Saturn"
planets are detected. In about 25\% of the cases where the ``Saturn"
planet is detected, the Jupiter planet is also detected. This is a
consequence of the fact that Saturn's orbital semi-major axis is only a
factor of 1.8 larger than Jupiter's orbital semi-major axis.
Such orbits are stable only if they are close to circular, so a space-based
microlensing survey will
be able to provide information on the abundance of giant planets with
nearly circular orbits by measuring the frequency of double planet detections
and the ratios of their separations. This is important information as giant
planets in Jupiter or Saturn-like orbits are thought to be required for
the delivery of volatiles, such as water, to the inner planets in the
habitable zone \cite{lunine}.

Events in which a terrestrial planet and a ``Jupiter" are
detected, such as the event shown in Figure \ref{fig-lc2}(b) are more
rare. In part, this is because the lower mass of the terrestrial planet
means that less of them will be detected, but another factor of is that
the ratio of Jupiter's semi-major axis to that of the terrestrial planets
is a factor of 3.5-7 rather than the factor of 1.8 ratio between the
Jupiter and Saturn orbital distances. Because of this, only 10-15\% of
the detected terrestrial planets will also have a Jupiter detection.

The detection sensitivity for multiple planets depends more on the
telescope size and the assumed level of systematic photometry errors
than the sensitivity for single planets does. More sensitive photometry
increases the probability that a planet can be detected in the light
curve of a microlensing event, and the number of double planets
detected depends on this probability squared. 

In order to estimate the sensitivity for detecting multiple planets,
we have calculated the detection probabilities for lenses with planetary
systems with the same planetary mass fractions and separations as the
planets of our own Solar System.  For the parameters of the
proposed GEST mission, we find that a total of about $150f$ multiple
planet systems will be discovered, where $f$ is the fraction of
planetary systems that resemble our own. About $13f$ of these will
be terrestrial-giant planet pairs, and the remainder will be multiple
planet detections consisting of only giant planets. (These numbers
assume that a lower detection threshold of $9\sigma$ can be used for the
second planets to be detected because there is a much smaller number of
light curves that must be searched for multiple planets.) A substantial
improvement in sensitivity can be obtained with the parameters of the
STEP mission: a 1.5m telescope with a 2.2 square degree field-of-view.
Half of the focal plane would use Lincoln Labs near-IR optimized CCD detectors
and the other half would use the HgCdTe IR arrays with a $1.7\mu$m 
cutoff. If we assume that the photometry is limited by systematic
errors of 0.15\% in a 10 minute exposure, then our simulation indicate
a total of $490f$ multiple planet detections with $45f$ of these
being terrestrial-giant planet pairs. 

\begin{figure}
\plottwo{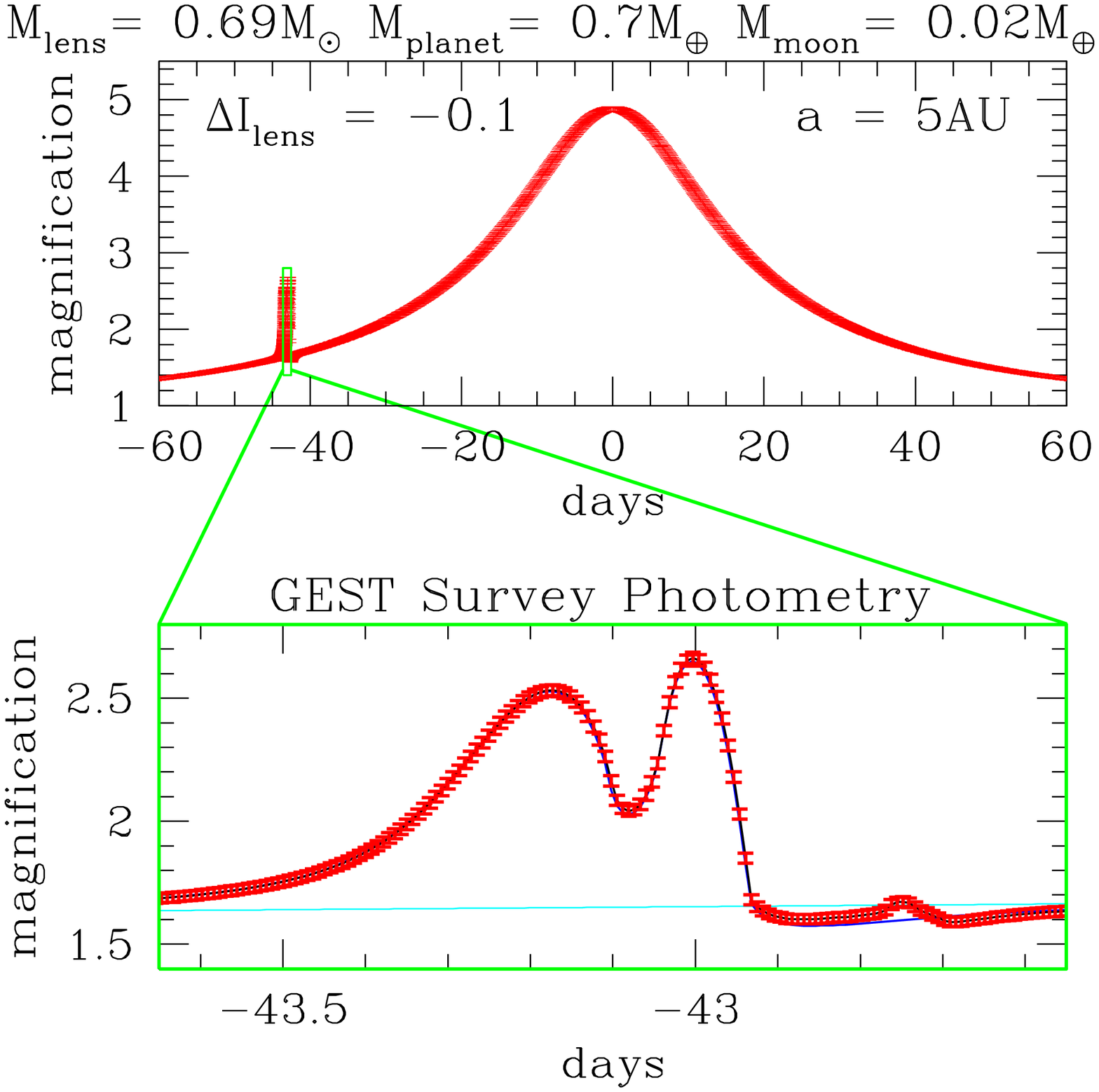}{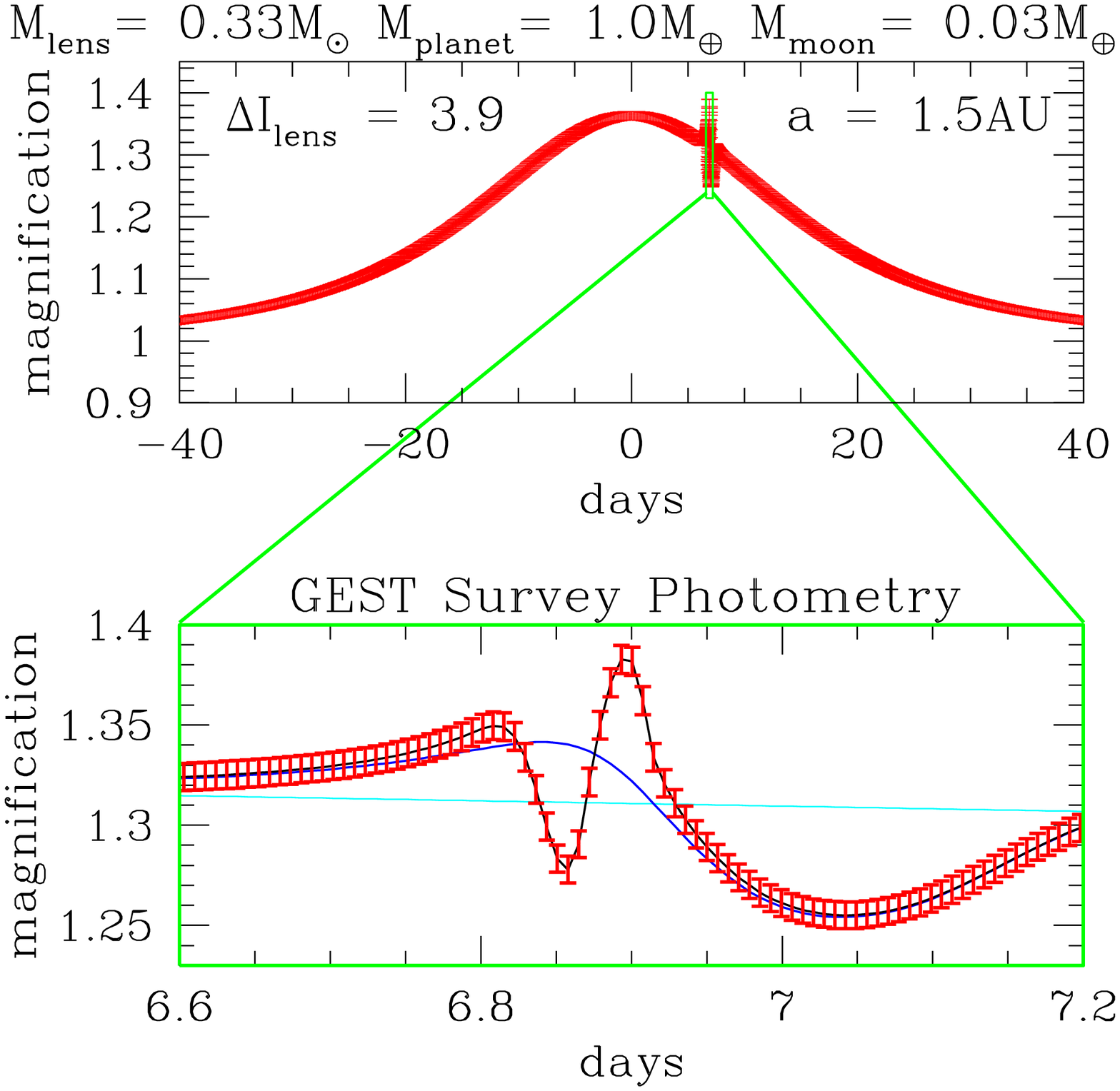}
\caption{
Example light curves of terrestrial planets with moons which have 1-2 times 
the mass of the Earth's moon. These moons orbit at $3.3$ and $0.56$ times
the Earth-moon separation, respectively.
\label{fig-lc_moon}}
\end{figure}

It is also possible to detect the large moons of terrestrial planets as 
shown in Fig.~\ref{fig-lc_moon}. The semi-major axis of the moon's 
orbit is about 0.8 times the Earth's Einstein radius, so systems like 
our own should be detectable. Because the planet-moon separation is
likely to be similar to the planetary Einstein radius, the light curve
deviations due to the planet and moon are likely to be closely spaced
in time or even overlapping as in the example shown in 
Fig.~\ref{fig-lc_moon}(b). Nevertheless, most of the light curve deviations
due to planet$+$moon systems are well approximated by the sum of the
deviations due the two minor masses by themselves. A more systematic
study of the detection of planet plus moon systems by microlensing
will be carried out in a future paper.

\subsection{Planet Detection Sensitivity}
\label{subsec-nplan}

\begin{figure}
\plotone{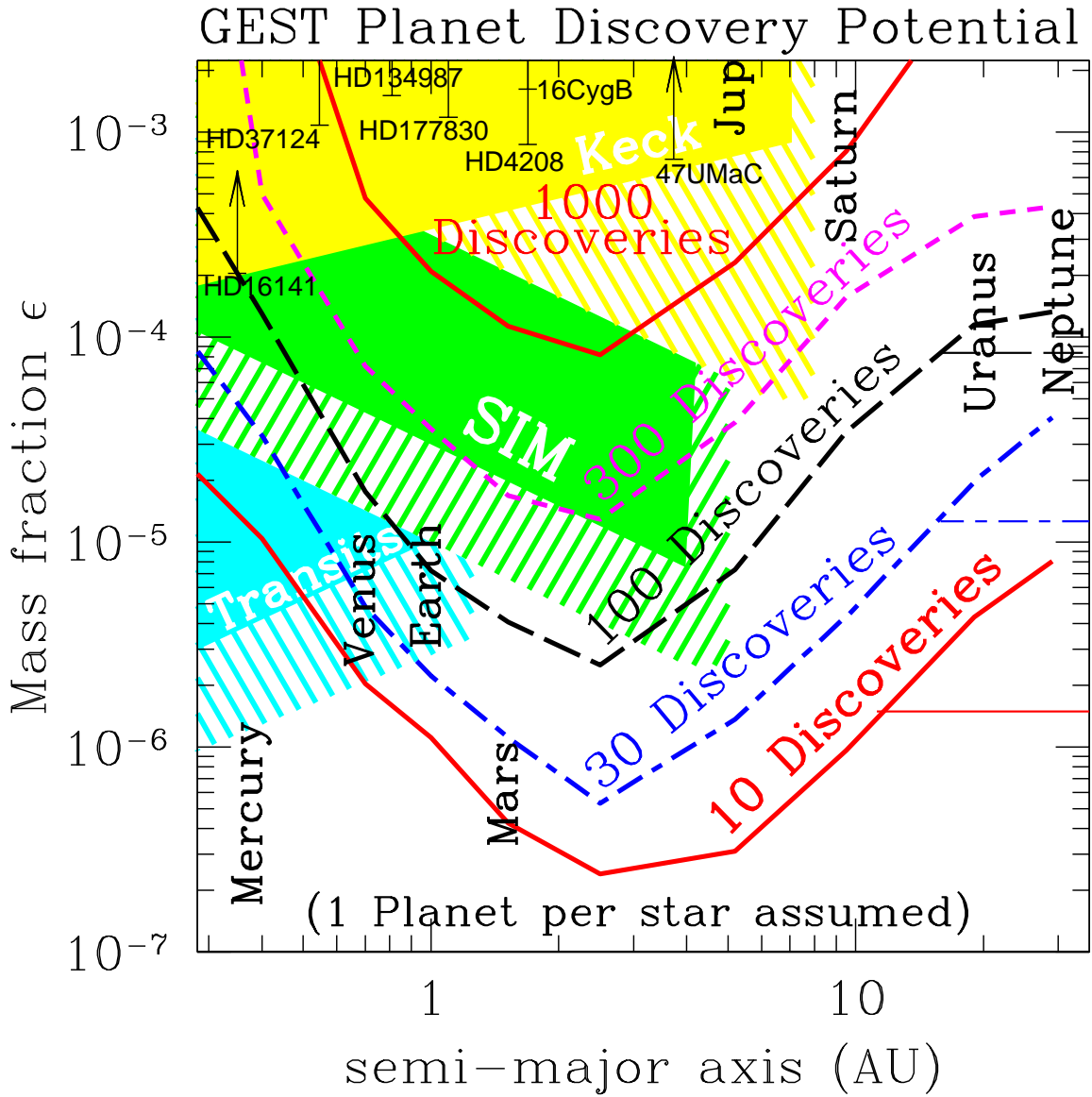}
\vspace*{-1.0cm}
\caption{
The sensitivity of the proposed GEST mission
is plotted as a function of planetary mass 
fraction, $\epsilon$, and orbital semi-major axis. 
The curves are contours indicating
the expected number of GEST planet discoveries assuming 1 planet per
star with the given parameters.
The solid yellow region gives the sensitivity of a 20-year radial 
velocity program on the Keck Telescope assuming a detection threshold of
10 m/sec, and the yellow lines indicate the 
sensitivity of a 10-year interferometric astrometry program with a
30$\,\mu$as detection threshold.
The green regions indicate the sensitivity of the
SIM recommended and floor missions. The location of our Solar 
System's planets and some of the extra-solar planets detected
by radial velocities are shown.
Most detected Earth mass planets 
have $\epsilon \approx 10^{-5}$ because the typical lens star has a mass
of $\sim 0.3 M_\odot$, so the plot indicates that GEST can see $\sim 35$ 
Earth-mass ratio planets at 1 AU and $\sim100$ Earth-mass planets at that 
distance. The horizontal lines indicate the sensitivity to free-floating 
planets since the more distant planets can sometimes be detected without 
seeing a microlensing signal from their star. 
\label{fig-n_vs_sep}}
\end{figure}

\begin{figure}
\plotone{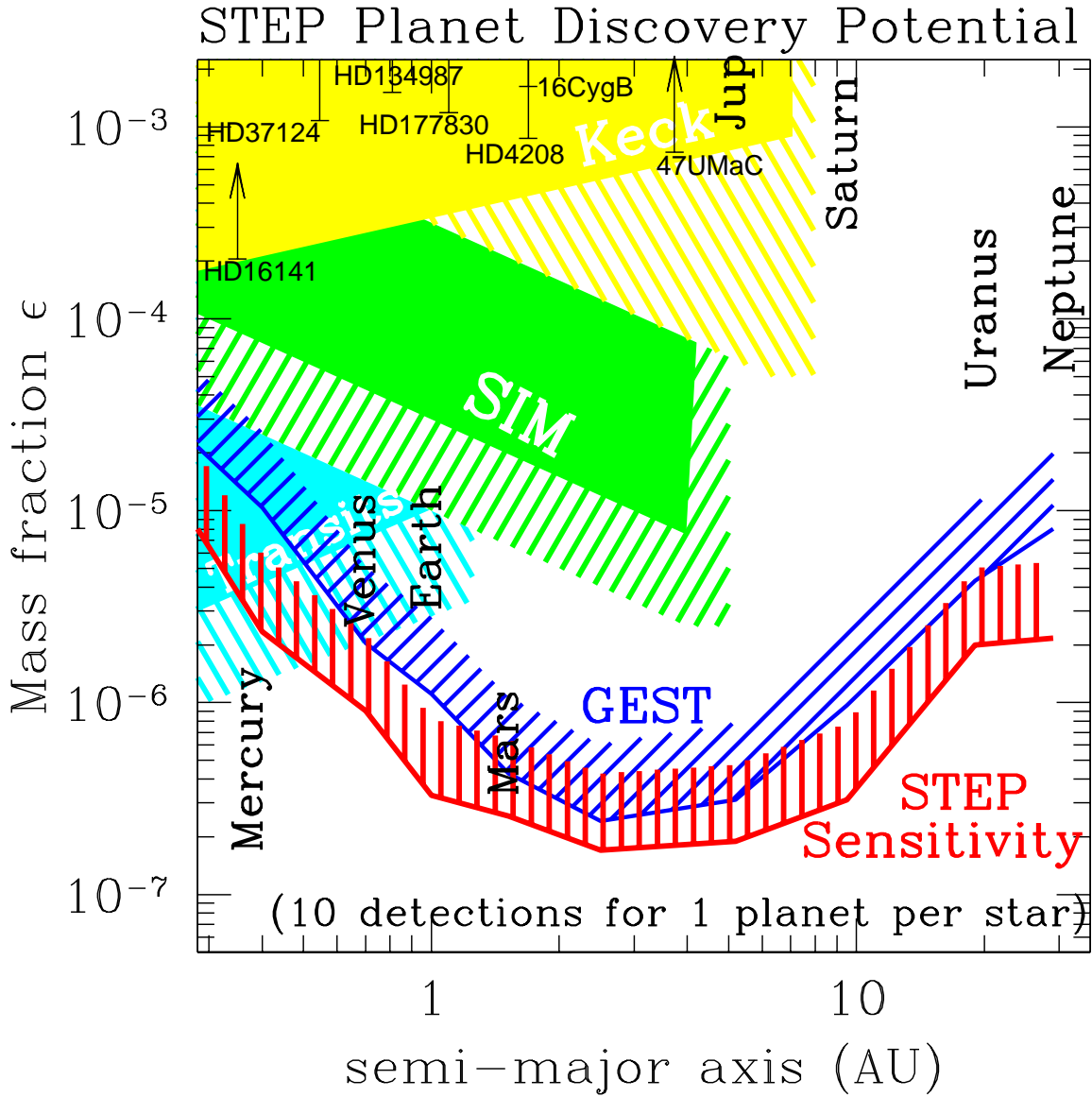}
\caption{
The sensitivity of the more ambitious STEP
mission sensitivity is compared to the sensitivity of the proposed
GEST mission, as well as the other planet search techniques shown
in Fig.~\ref{fig-n_vs_sep}. The improvement in sensitivity due to
STEP's larger telescope and more sensitive detectors is more pronounced
at large and small separations than at the region of maximum sensitivity
at 2-3$\,$AU. This is because the more sensitive mission is able to
detect planetary signals of a smaller amplitude which often occur for
planets with separations that are significantly smaller or larger than 
the Einstein ring radius.
\label{fig-n_vs_sep_comp}}
\end{figure}

\begin{figure}
\plotone{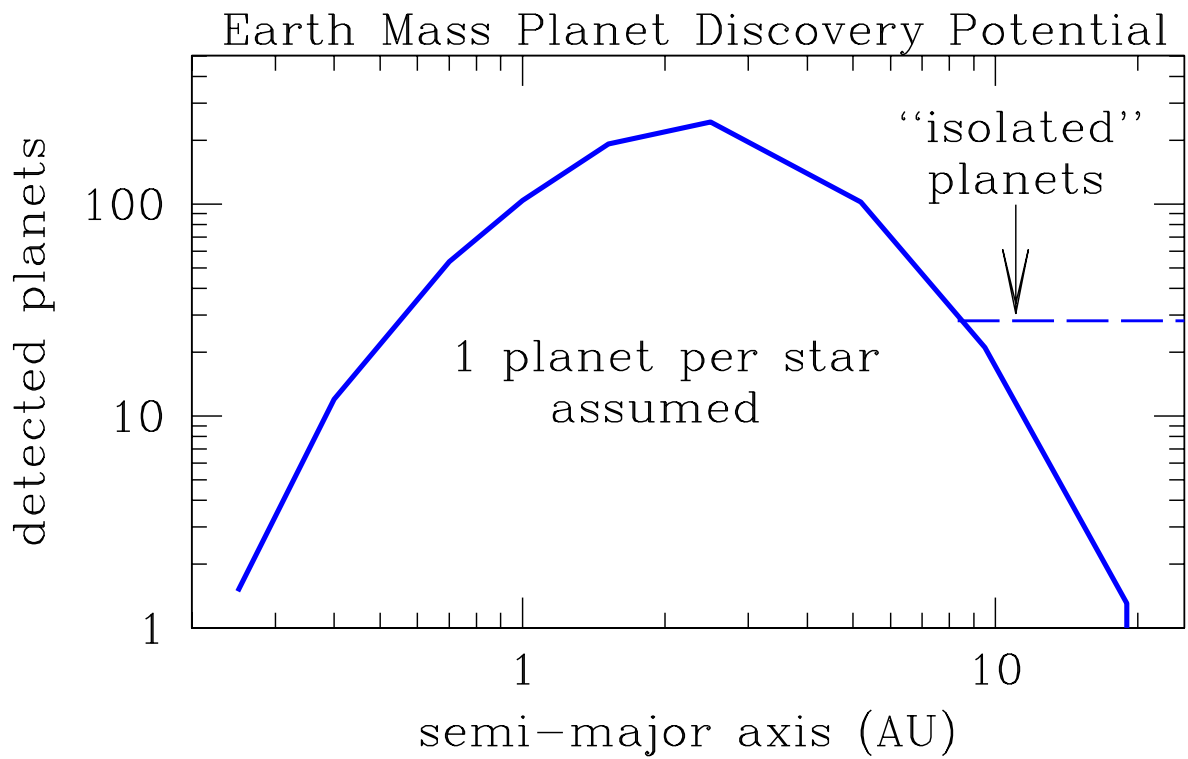}
\caption{
This is a plot of GEST's sensitivity to Earth-mass planets. The number
of detected Earth-mass planets is shown as a function of the
orbital semi-major axis assuming one such planet per lens star.
At a semi-major axis of $\sim 10\,$AU, the number of planet detections
reaches the lower limit of about 30 set by the free-floating planet
detection calculation. Most of the planets detected with 
semi-major axis $\gg 10\,$AU will be detected in ``isolation\rlap," without
a detection of their host star.
\label{fig-earthmass}}
\end{figure}

The major goal of our simulations is to determine
the sensitivity of a space-based microlensing survey. The 
sensitivity to planets orbiting each 
of the lens stars depends on a large number 
of factors including the event timescale, the size of the photometric 
error bars, and the angular size of the source star. Thus, the simplest
way to display the planet detection sensitivity to is
to give the number of expected planet detections under the assumption
that each lens star has a
planet of a given mass fraction, $\epsilon$, and separation. This is what is
plotted in Figs.~\ref{fig-n_vs_sep} and \ref{fig-n_vs_sep_comp}.
The different curves in Fig.~\ref{fig-n_vs_sep} are
contours of constant numbers of planet discoveries, assuming one planet 
per star at the given mass fraction and semi-major axis. In 
Fig.~\ref{fig-n_vs_sep_comp}, we compare the sensitivities of
the proposed GEST and STEP missions.  The locations of the
planets in our Solar System are also shown. Each planet name starts at the
planetary mass fraction of the planet and continues toward higher mass
fractions. Because the typical mass of a lens star is about $0.3\,\msun$,
planets of the same mass as the Solar System's planets will have a typical
mass fraction that is larger by about a factor of three. A planet of one 
Earth mass, for example, will usually have $\epsilon \approx 10^{-5}$ 
rather than $\epsilon = 3\times 10^{-6}$, which is the Earth's mass
fraction. So, the sensitivity to planets with the same mass as
those in the Solar System will appear near the top of each planet name
while the bottom of each planet name indicates the sensitivity to planets
of a fixed mass fraction. The sensitivity to planets of $1\,\mearth$ 
for the parameters of the GEST mission is
shown in Figure \ref{fig-earthmass} which indicates that just over 100 Earths
would be detected if each lens star has one in a $1\,$AU orbit. The peak
sensitivity is at an orbital distance of $2.5\,$AU where we would expect
230 detections if each lens star star had a planet in such an orbit.

The green and yellow shaded regions in Figure \ref{fig-n_vs_sep} indicate
the sensitivity of other planet search techniques. The known extra-solar
planets which orbit main sequence stars have been discovered with the
precision radial velocity technique \cite{marcy-but}, and a number of
these individual detections are indicated in the upper left region of
the figure at small semi-major axes and large masses. The solid yellow
shaded region indicates the sensitivity of a 20-year radial velocity 
program assuming a minimum detectable velocity amplitude of $10 {\rm m/sec}$.
This is close to the demonstrated accuracy of the 
Keck \cite{marcy-but} and CORALIE \cite{coralie}
radial velocity programs, but it is expected that the
current radial velocity state of the art is close to the limit set
by the intrinsic radial velocity noise of the source stars.
The expected sensitivity of the planned 5-year
Space Interferometry Mission (SIM) satellite is shown in green with the
vertical green lines showing the planned SIM sensitivity and the solid
green region showing the sensitivity of the SIM floor mission. (The 
assumed detectable astrometric signals are $1\,\mu $as and $6\,\mu $as,
respectively, at a distance of $10\,$pc.)

The cyan shaded region in Figure \ref{fig-n_vs_sep} represents the
space-based transit technique which is very sensitive to terrestrial planets
in short period orbits. Several such transit missions are planned including the
French COROT mission, ESA's (not yet funded) Eddington mission,
and NASA's Kepler mission. Kepler will be the most sensitive of these,
and its sensitivity is represented by the diagonal cyan lines.
A sensitive transit search like Kepler is the only program that is
competitive with a space-based microlensing survey for finding
Earth-mass planets at $1\,$AU. However, the prime sensitivity of a transit
survey extends inwards from $1\,$AU, while the sensitivity of microlensing
extends outwards. So, the two methods are largely complementary.

Figs.~\ref{fig-n_vs_sep} indicates that microlensing's peak planet detection
sensitivity is 
at 2-3$\,$AU with significant sensitivity in the range 0.7-10$\,$AU.
In fact, the sensitivity at large distances is underestimated by our
simulation because we do not consider planets that may be detected
when the source star magnification is $A < 1.06$. Events with
$A_{\rm max} < 1.06$ and events with the planetary deviation which
occurs before or after the $A > 1.06$ region of the light curve have not been
included in our simulations. However, some of these planets will be
detectable. A lower limit on our sensitivity to distant planets
is set by our sensitivity to free-floating planets which is discussed
in section \ref{subsec-ffloat}. This sensitivity is indicated by the
thinner, horizontal lines on the right side of Figure \ref{fig-n_vs_sep}.
These lines should be considered to extend to infinite distances,
indicating that a space-based microlensing survey has strong sensitivity to
planets at separations of $0.7\,$AU to $\infty$.
However, for planets at distances $\gg 10\,$AU, it will often be the case
that the star that the planet orbits will not be detectable. Such cases
may be difficult to distinguish from free-floating planet detections
unless the lens star can be detected 
(see Section \ref{subsec-sstar}).

Microlensing of Galactic bulge stars is most sensitive at semi-major axes of
2-3$\,$AU because this is the typical Einstein ring radius for 
Galactic bulge source stars. Images are located close to the Einstein ring
when they are bright, and the a planet is most easily detectable if
one of the bright images passes close to it. In contrast, the 
astrometry technique is more sensitive at large orbital radii, while the
radial velocity and transit techniques (see section \ref{subsec-transit})
are more sensitive at smaller radii. The astrometry, radial velocity,
and transit techniques all have sharp cutoffs on their sensitivity
at larger semi-major axes due to the fact that these techniques
require data from a full orbit, or several orbits in the case of
transits.
Thus, microlensing has an advantage over these other techniques
at large orbital distances, since it is able to make prompt discoveries of
distant planets.

The main advantage of the microlensing technique over both the astrometry
and radial velocity techniques is its sensitivity to lower mass planets. 
At $1\,$AU, microlensing is sensitive to planets with masses that are
about three orders of magnitude smaller than the smallest masses that 
ground based radial velocity and astrometry searches are likely to detect.
A space based microlensing survey also offers an advantage in sensitivity
to low mass planets with respect to space based astrometry missions
such as SIM. Figure \ref{fig-n_vs_sep} indicates that GEST's sensitivity
extends to masses that are a factor of 20 lower than expected for the
SIM baseline mission and a factor of 100 lower than for the SIM floor
mission. (The floor mission is considered to be
the minimum acceptable sensitivity 
that SIM could descope to if it should run into budget problems.)
Of course, SIM will
find planets orbiting nearby stars, so planetary results to be expected from
the GEST and SIM missions are somewhat complementary: GEST will determine
extra-solar planet abundances extending down to very low masses, while
SIM will study planetary systems close to the Sun with sensitivity down
to planets somewhat more massive than the Earth.

\begin{figure}
\plotone{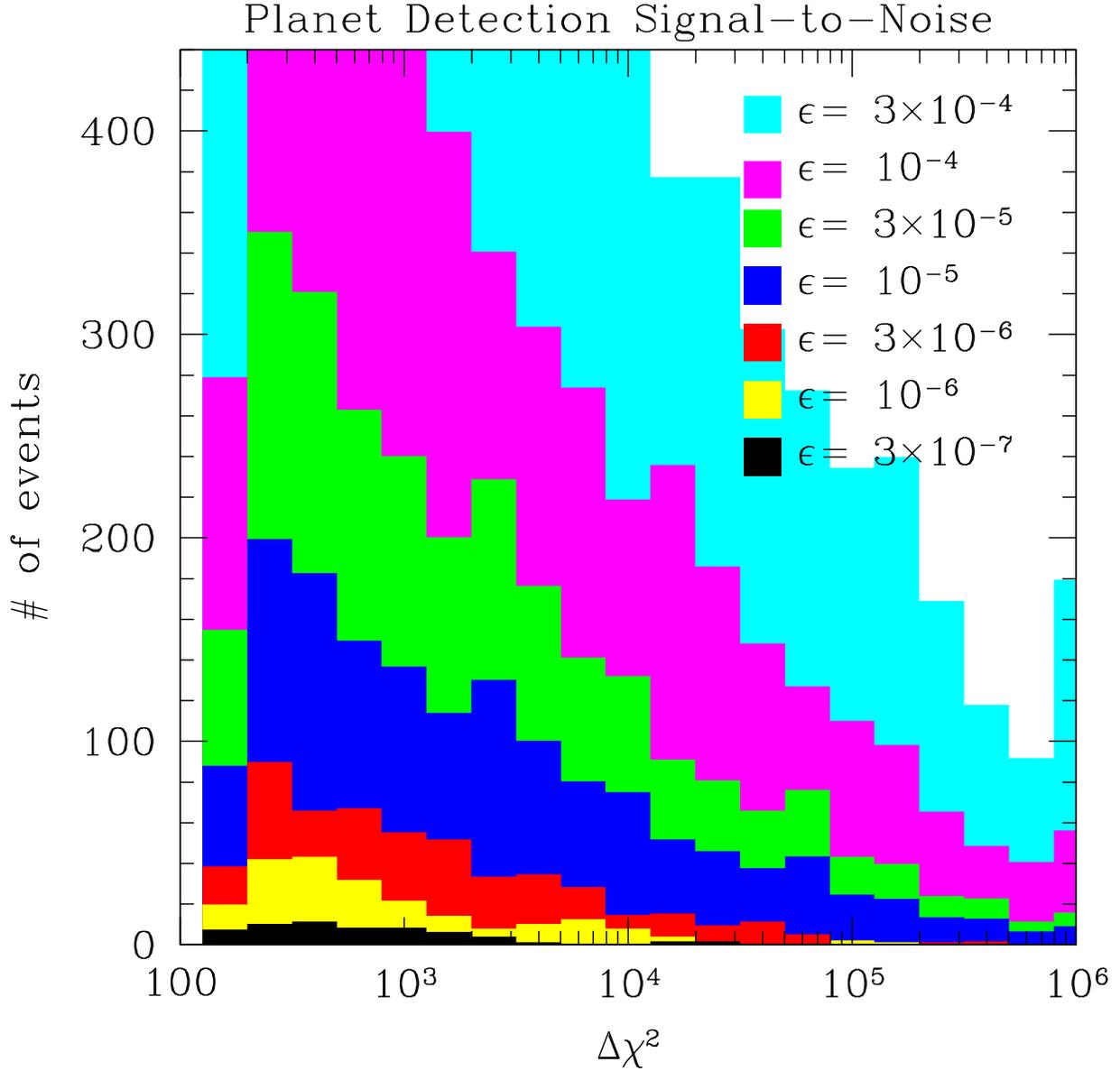}
\caption{
This is a histogram of the planetary detection significance, $\Delta\chi^2$,
for different mass fractions, $\epsilon$, ranging from 
$\epsilon = 3 \times 10^{-7}$ (the mass fraction of Mars) to 
$\epsilon = 3 \times 10^{-4}$ (the mass fraction of Saturn). 
For planets with an Earth-like mass fraction ($\epsilon = 3 \times 10^{-6}$) 
and above, more than half of the detected events have $\Delta\chi^2 > 800$ 
which corresponds to a $28\sigma$ detection.
\label{fig-signoise}}
\end{figure}

Another important advantage of the gravitational microlensing technique
is that the low mass planets are detected with high signal-to-noise.
In fact, for a large range of planetary masses, the strength of the
microlensing signal does not depend on the mass of the planet. Low mass
planets do affect a smaller region of the lens plane, so they have
a lower detection probability and a shorter duration. Figure \ref{fig-signoise}
shows the distribution of the signal-to-noise of our detected planets
for planetary mass fractions ranging from $\epsilon = 3\times 10^{-7}$ 
(Mars-like) to $\epsilon = 3\times 10^{-4}$ (Saturn-like). $\Delta\chi^2$
is the detection significance parameter used for the x-axis of this
plot, and a logarithmic scale must be used because of the large
spread in $\Delta\chi^2$ values. The most striking feature of this figure
is that number of events with large $\Delta\chi^2$ values falls off
rather slowly. The power law, $N \sim \left(\Delta\chi^2\right)^{-1.3}$, 
provides a rough fit to these curves for all but the lowest mass fraction
($\epsilon = 3 \times 10^{-7}$) where the effects of the finite angular
size of the source stars begin to reduce the number of high
signal-to-noise events.

\subsection{Sensitivity Dependence on Telescope Parameters}
\label{subsec-depend}

Tables 1 and 2 summarize how the planet detection sensitivity for Earth-like
planets depends on the parameters of the space-based microlensing
survey telescope. The parameters varied are the telescope field of view,
the assumed minimum photometric error in a 10 minute exposure, the
assumed FWHM of the images, and the effective telescope aperture in meters.
The FWHM and aperture are considered independently because they
can be varied independently when the pass-band and telescope optics
are varied. The pass-band and detector sensitivity contribute to the
effective aperture by modifying the total number of photons detected.
The effective aperture is normalized assuming the detector
quantum efficiency of the standard CCDs shown in Fig.~\ref{fig-qe}
with a broad 0.5-$0.9\,\mu$ pass-band and a telescope optical
thruput of 70\%. Narrower pass-bands can decrease the effective
aperture, and the use of more sensitive detectors can increase
the effective aperture. Thus, the telescope proposed for the
GEST MIDEX proposal has an effective aperture of 1.25m even though
the actual aperture is 1.0m because the more sensitive Lincoln
Labs CCDs are used with a 0.5-$1.0\,\mu$ pass-band.

The planet detection sensitivity has a weaker dependence on a number of
these parameters than might naively be expected. For example, the number of
planets detected does not depend linearly on the field-of-view because
we are able to select a field with a higher average microlensing optical
depth when the field is smaller. Also, the dependence on the image FWHM is
relatively weak because all of the values considered allow stars near
the top of the bulge main sequence to be individually resolved. The
sensitivity decreases quite substantially at FWHM $\simgt 0.5$", however. 

We should caution that the main advantage of a more sensitive
telescope, like the proposed STEP mission, is the increased sensitivity to
multiple planet detections. As described in sub-section \ref{subsec-lcmult},
the proposed STEP mission should expect to detect 3-3.5 times more 
multiple planet systems than the proposed GEST mission would. Some of
this increase in sensitivity to multiple planets is due to the fact that
the probability of detecting two planets scales like the single planet
detection probability squared. However, when one planet is detected, it 
often has a separation that is close to the Einstein ring radius. Since
a second planet is likely to have a separation that is not close to the
Einstein ring radius, it will likely have a weaker than average signal. Thus, 
the ability to detect multiple planets is more sensitive to the
telescope size and detector sensitivity than the square of the single
planet detection probability.

\subsection{Variable Star Background}
\label{subsec-vars}

All of our simulations have implicitly assumed that there is no
significant background of variable stars that might interfere with the
detection of planets. Some justification for this is provided by the
existing gravitational microlensing surveys which have not seen a significant
background of variable stars 
(\citeNP{macho-bulge45,macho-bulge-diff,ogle-microcat}). In fact,
the most significant source of variability that might contaminate
samples of gravitational microlensing events is
background supernovae \cite{macho-lmc5.7}. However, the space-based
microlensing program that we propose will use source stars that are
fainter than the source stars used for the ground-based surveys. Faint
flare stars \cite{flare} are of particular concern because they can
have long quiescent phases with infrequent brightenings seen in
broad-band photometry. However, this broad-band variability is 
generally seen in the blue or ultra-violet bands, and is much less
pronounced in the red and near-IR where microlensing surveys would
observe. 

While the ground-based microlensing surveys follow relatively bright stars
in the Galactic bulge and Large Magellanic Cloud, they also observe many
thousands of intrinsicly fainter stars in the foreground of these targets.
None of the foreground stars observed by the MACHO Collaboration has exhibited
the sort of photometric variation that could be confused with a planetary
microlensing deviation if, by chance, the intrinsic stellar photometric
variation occurred during a stellar microlensing event. Since we expect
about $10^4$ stellar microlensing events, the statistics of the foreground
stars observed by the ground-based surveys suggest that there should be
no contamination of the planet sample due to variable star.  The data provided
by a space based survey will provide much more stringent constraints on
possible variable star contamination, and we expect that the accurate
measurements of the light curve shape from a space-based survey will clearly
distinguish between deviations due to microlensing and any intrinsic 
variability of the source star. It is likely that the variable star
background will have a negligible effect on the sensitivity of a space-based
gravitational microlensing planet search program.

\subsection{Free Floating Planets}
\label{subsec-ffloat}

The leading theories of planet formation (\citeNP{levison,perryman}) 
indicate that
planets often don't stay in the same orbit where they formed. The migration
of giant planets inward is thought to be necessary to explain the
``hot Jupiter" planets discovered by the radial velocity planet searches,
and the orbital distribution of Kuiper Belt Objects \cite{malhotra}
suggests that Neptune has migrated outward from its birth site.
These migrations are likely to be due to the gravitational interactions
of these giant planets with a large number of planetesimals in the
protoplanetary disk. Many of these planetesimals are likely to be perturbed
into highly elliptical orbits which will send them crashing into the Sun
or ejecting them from the solar system, and it is expected that the
most massive of these ejected objects will have a mass in the
terrestrial planet range which means that they should be detectable via
microlensing.

\begin{figure}
\plotone{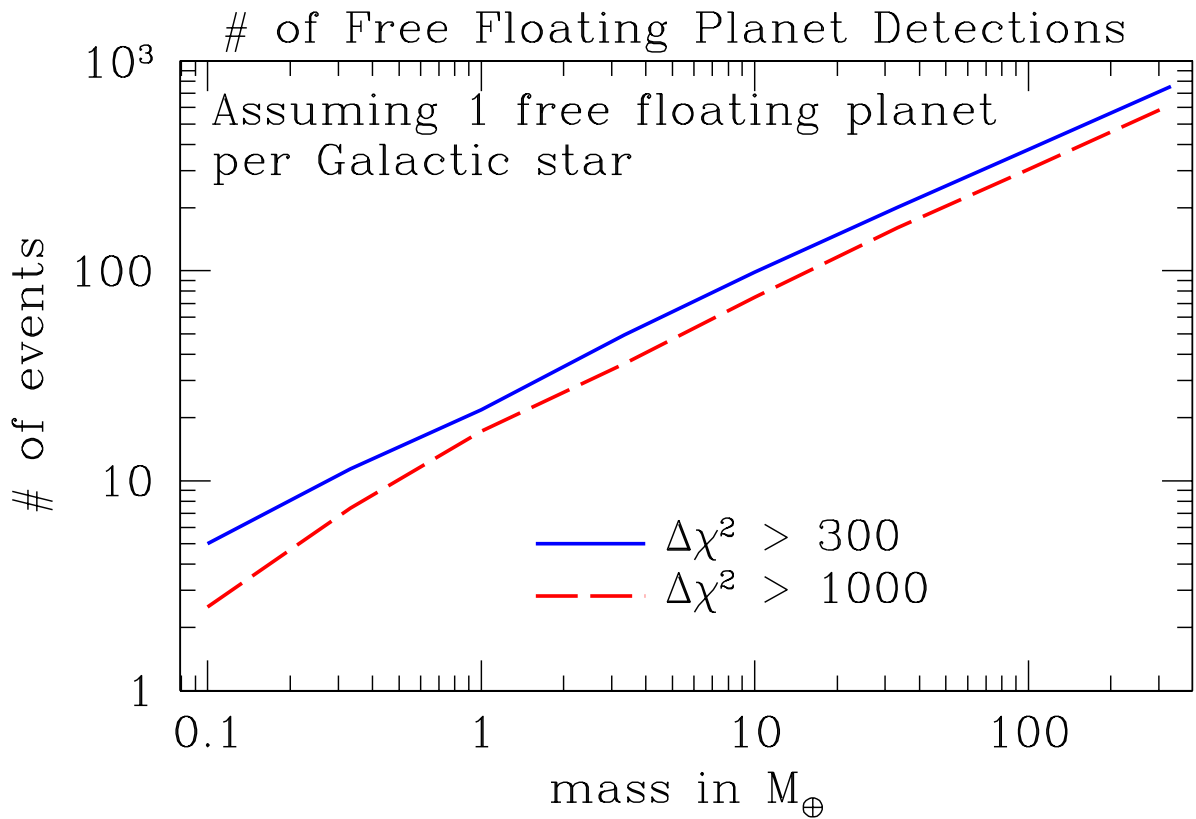}
\caption{
The number of free-floating planets to be discovered by GEST vs.~planetary 
mass for 2 different detection criteria which are equivalent 
to $17\sigma$ and $30\sigma$, respectively.
\label{fig-nff}}
\end{figure}

The majority of known extra-solar giant planets in orbits of semi-major
axis $> 0.3\,$AU have relatively large orbital eccentricities, and this
can be explained  via gravitational scattering with other giant planets
in the same system \cite{levison}. A consequence of these interactions is
that many of these giant planets will be ejected from their planetary
system. Terrestrial planets, which are more easily ejected via two-body
interactions, should also be ejected in large numbers.
Thus, there are good theoretical reasons to believe that 
free-floating planets may be abundant as a by-product of the planetary
formation process. If so, they can be detected via gravitational
microlensing. Figure \ref{fig-nff} shows the number of free-floating
planet detections expected for the GEST mission under the assumption that
there is one free-floating planet per Galactic star. The detection
threshold is set higher for the free-floating planet detections because
we must search $\sim 10^8$ light curves for free-floating planets
while we only need to search the $\sim 10^4$ detected stellar microlensing
event light curves for evidence of bound planets.  Since theory predicts
that many stars may be ejected from the system during the planetary
formation process, it may be reasonable to assume that there will
be many more free-floating planets than the numbers indicated in
Figure \ref{fig-nff}. If half of the star systems eject an average of
ten $1\mearth$ planets each, then we would expect to detect more than 100.
In fact, there has already been a possible detection of
a free-floating planet in the MACHO data \cite{macho_planets}.

\subsection{Source Star Identification}
\label{subsec-sstar}

\begin{figure}
\plotone{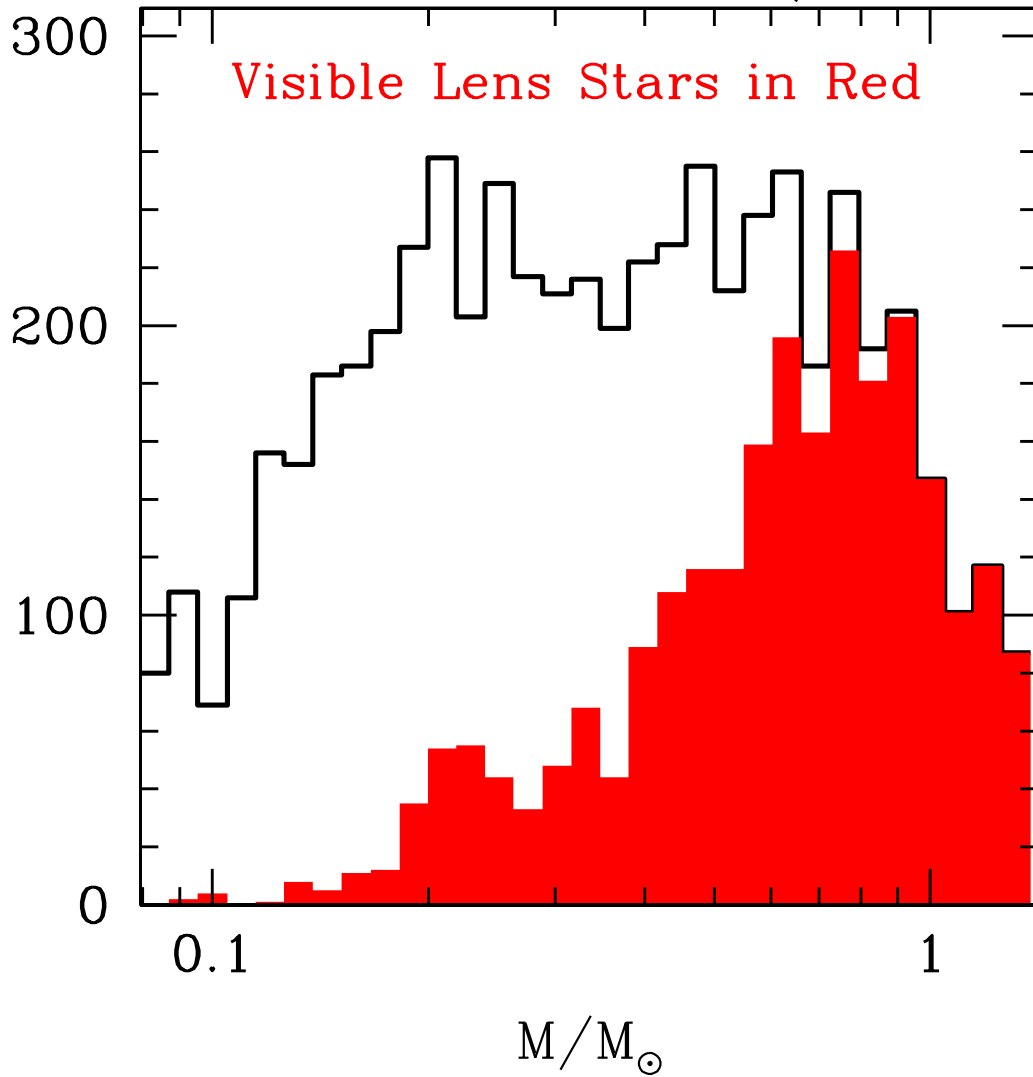}
\caption{
The simulated distribution of stellar masses is shown for stars 
with detected terrestrial planets. Lens stars are considered visible 
when they are at least 10\% of the brightness of the source star, if
they are not blended with a brighter star (besides the source). 
1/3 of the events have visible lens stars.
\label{fig-lens_detect}}
\end{figure}

The planets detected by and space-based microlensing survey 
orbit the lens stars in the foreground of the Galactic bulge source 
stars. The mass distribution of the lens stars from our GEST 
simulations is shown in Fig.~\ref{fig-lens_detect}. 
This distribution is somewhat 
flatter than the stellar mass function because we have assumed that 
the planetary mass distribution is proportional to the stellar mass 
distribution and more massive planets have a higher detection probability.

Although microlensing does not require the 
detection of any light from the lens stars, a 
significant fraction of the microlensing events seen 
by a space-based microlensing survey
will have lens stars that are bright enough 
to be detected. Our simulations indicate that for 
$\sim 17$\% of the detected planets, the planetary host 
(lens) star is brighter than the source star, and for 
another $\sim 23$\% the lens stars that is within 2.5 I-band 
magnitudes of the source star's brightness. A few of 
these stars are blended with the images of other 
brighter stars, and if we ignore those stars, we find 
that 33\% of the lens stars should be directly 
detectable. The detectable planetary host stars are 
depicted in red in Fig.~\ref{fig-lens_detect}, and they comprise 
virtually all of the F and G star lenses, most of the K 
star lenses, and a few of the nearby M-star lenses.

The visibility of the lens star will allow for the 
measurement of a number of other useful 
parameters. The most obvious of these are the 
apparent magnitude and color of the lens star. This 
would enable an approximate determination of the 
lens mass and distance if the dust extinction was 
small. Our field, however, has high and variable 
extinction, and so it will be prudent to obtain IR 
photometry. This will allow us to estimate both the 
extinction and the intrinsic color of the star. Because 
our fields are quite crowded, we will need IR 
observations with high angular resolution which can 
be obtained with adaptive optics (AO) systems on 
large telescope such as the VLT, Gemini, LBT or 
Keck. The high stellar density of the microlensing survey fields
implies that there are virtually guaranteed to be nearby guide stars 
to provide the phase reference needed for 
these AO systems. We would expect to obtain two 
sets of IR, AO observations: one during the event 
which would be scheduled as soon as the planetary 
signal is detected and the second set of observations 
would be taken well after the event is over. This pair 
of observations taken at different lens magnifications 
will allow us to unambiguously determine the color 
and brightness of the lens stars. We will require this 
data only for events with detected planetary signals, 
and so there should no difficulty in obtaining the 
ground-based telescope time.

Another measurable parameter for the visible lens 
stars is the relative proper motion between the lens 
and the source which is typically $\mu\approx 8\,$mas/yr for a 
total motion of 32 mas over 4 years. This is 15\% of a CCD pixel
for the sampling of the proposed GEST mission. 
Anderson and King (2000) argue that centroids can 
be measured to 0.2\% of a pixel with a combination 
of a set of undersampled HST WFPC2 frames that have been
dithered to recover the resolution lost to undersampling. 
A space-based microlensing survey
will provide $>100$ times more data than the 
most ambitious HST programs, which will allow 
numerous cross-checks to look for systematic errors 
in the centroid determinations. Thus, we expect that 
the centroids of the space-based stellar images can be determined at 
least as well as the centroids of the HST stars, so we 
expect to be able to measure the relative proper 
motion to an accuracy of a few percent. An 
independent measurement of the lens-source proper 
motion can be obtained for the events which exhibit 
planetary lens caustic crossing features. These 
comprise somewhat more than 50\% of the events in 
which terrestrial planets are detected, and they allow 
the ratio of the angular radius of the star to the 
angular Einstein radius, $\theta_E$, to be measured in the 
light curve fit. Since the source star angular radius 
can be estimated from its brightness and color, an 
estimate of $\theta_E$ can be obtained. The ratio of the 
angular Einstein radius to the lens-source proper 
motion is $\theta_E/\mu = t_E$, the Einstein radius crossing time 
which can also be measured from the light curve, and 
so these measurements of $\mu$ and $\theta_E$ give equivalent 
information.

\begin{figure}
\plottwo{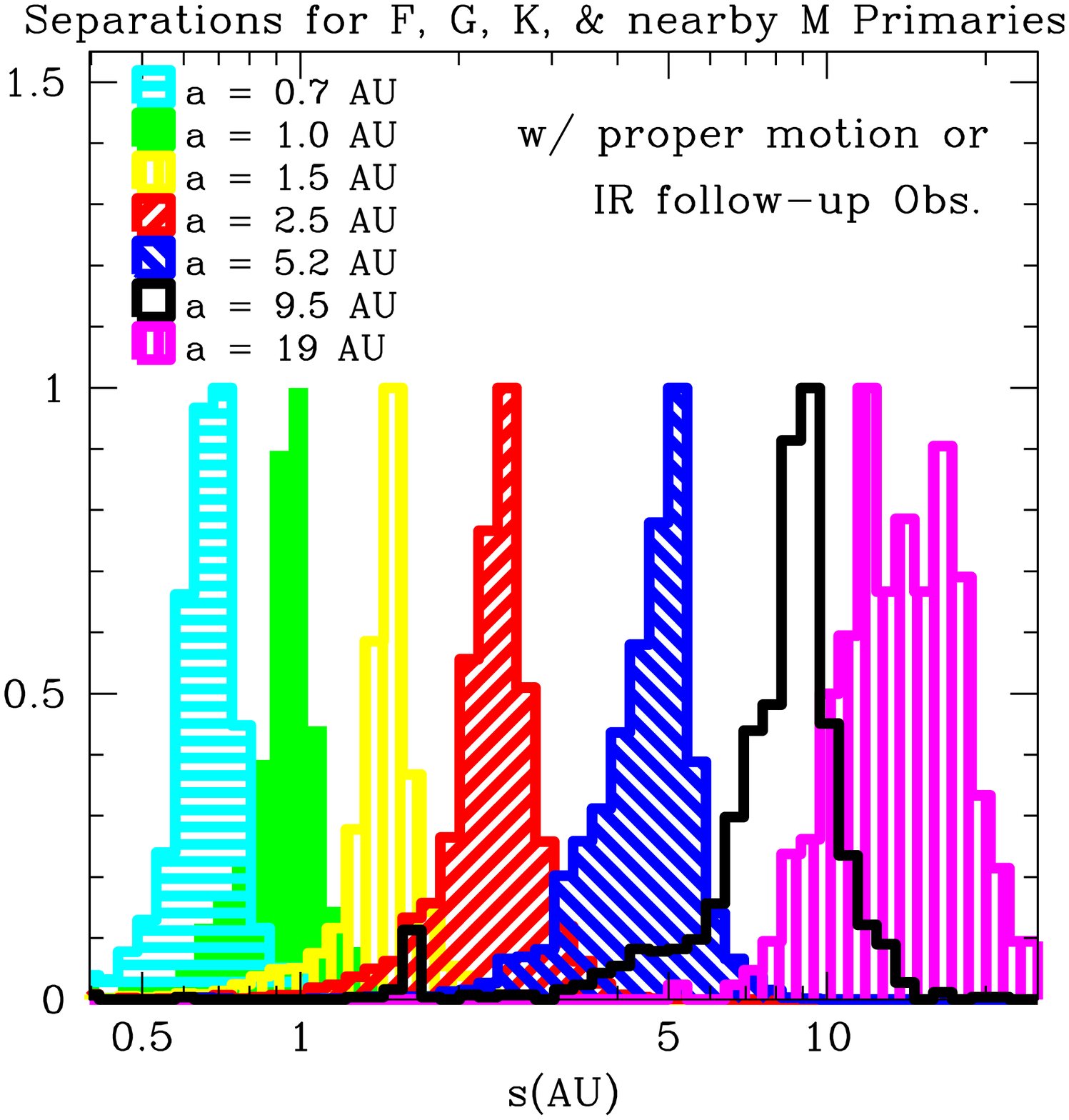}{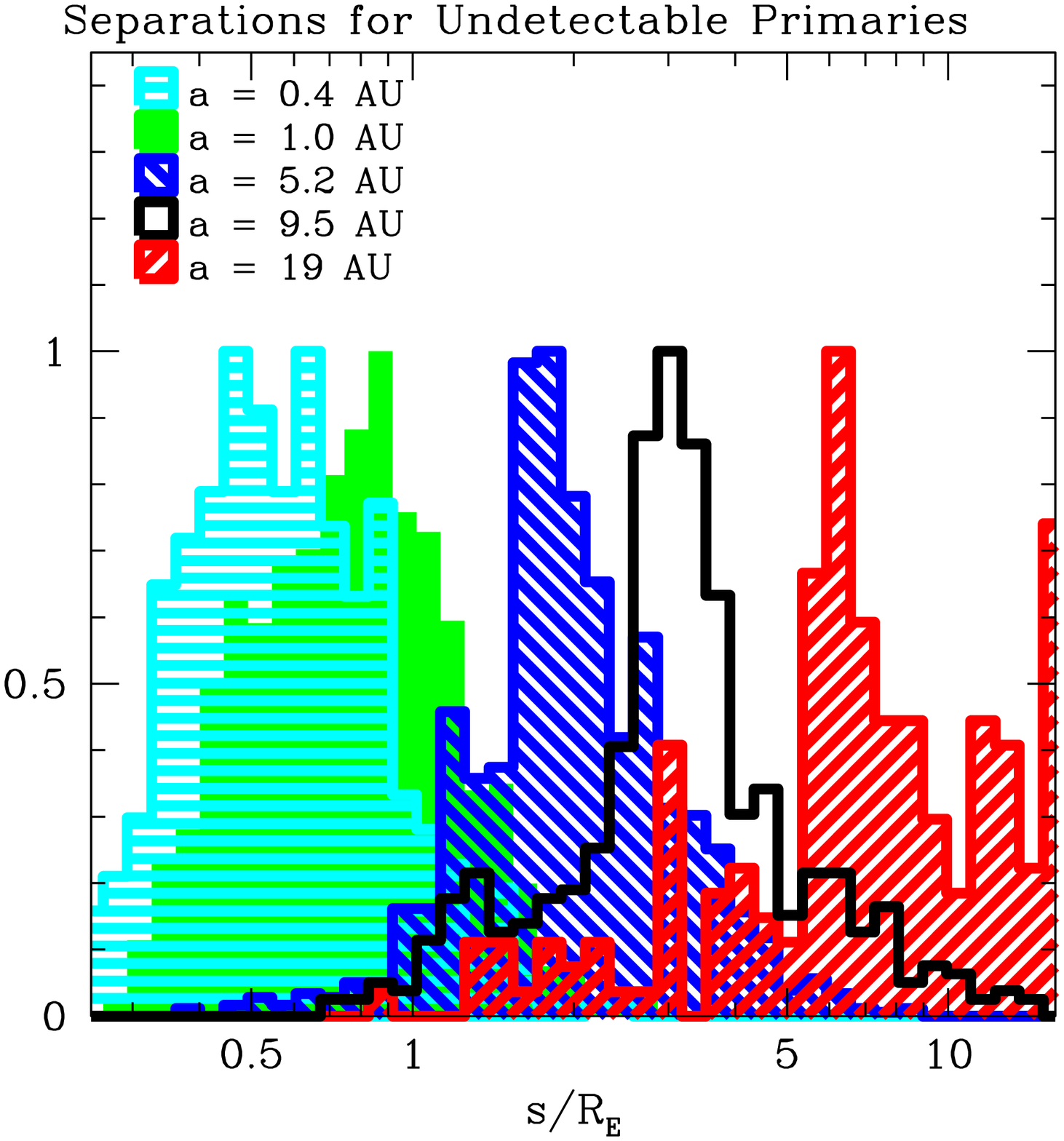}
\caption{
In the left panel, the
measured separation for planets with different orbital semi-major axes 
is shown for the ``visible" lens stars, for which the Einstein radius, $R_E$, 
can be determined . This allows the conversion of the
measured separation, s, into physical units (AU). The following 
measurement accuracies are assumed: lens
$I_{\rm lens}$: 20\%, $I-K_{\rm lens}$: 10\%, lens-source star centroids: 
2 mas. The observed scatter in the measured separation
relation is mostly due to the projection of the orbital plane on the sky.
The distribution of measured star-planet separations is shown on the
right for detected planets which orbit undetectable stars. The observed 
correlation between the planetary semi-major axis indicates that the
measured separation can be used to estimate the semi-major axis 
with an accuracy of a factor of 2-3.
\label{fig-sepdist}}
\end{figure}

The measurement of $\mu$ or $\theta_E$ allows us to use the 
following relation for the lens star mass, 
\begin{equation}
M_l = {\theta_E^2 D_s c^2 \over 4G} {x\over 1-x}
\end{equation}
where $x=D_l/D_s$, the ratio of the lens to source 
distances. This relation allows us to determine the 
difference between the source and lens distances 
when the lens is close to the source because it 
indicates that $M_l$, and hence the lens luminosity,
depends sensitively on $1-x = (D_s-D_l)/D_s$. This means 
that the Einstein radius, $R_E$, can be determined for all 
lens stars with a measurement of the lens star 
brightness and its relative proper motion, $\mu$, or its 
color, which in turn, implies that the planetary separation can 
be determined in physical units. The results of this 
determination are shown in Fig.~\ref{fig-sepdist}, which 
shows the measured separation for detected planets 
as a function of their orbital semi-major axis. For this 
plot we have assumed that the change in the relative 
lens-source centroid can be measured to 2 mas, the 
reddening corrected I magnitude of the lens can be 
measured to an accuracy of 0.2 mag., and the 
reddening corrected I-K color can be measured to 
0.1 mag. As Fig.~\ref{fig-sepdist}, indicates, the resulting 
estimate for planetary semi-major axis is accurate to 
about 20\%. The uncertainty is dominated by the 
unmeasured distance along the line-of-site.
When the lens star cannot be detected, the 
projected separation between the planet and its host 
star can only be measured in units of $R_E$. This can be 
used to estimate the planetary orbit semi-major axis 
by means of the expected correlation shown in the
right hand panel of Fig.~\ref{fig-sepdist} 
which indicates that physical separation can 
be estimated with an accuracy of a factor of 2 or 3.  

\subsection{Measurable Planetary Parameters}
\label{subsec-ppar}

The utility of planets that are detected by a space-based gravitational
microlensing survey depends, of course, on the planetary
properties that can be measured. For the $\sim 33$\% of events with
lens stars that are bright enough to be detected, the following parameters
can be measured:

\begin{itemize}
  \item The mass of the planetary host (and lens) star 
        is determined (with some redundancy) from the microlensing event
        time scale, the lens-source proper motion, $\mu$, and the source
        brightness and color.
  \item The planetary mass, $M_{\rm planet}$, is determined from the
        the stellar mass and planetary mass fraction, $\epsilon$, which
        comes from the microlensing light curve fit.
  \item The distance to the planetary host star is determined
        from the same combination of parameters that gives the stellar mass.
  \item The planet-star separation (in the plane of the sky) is always
        measured in units of the Einstein ring radius, $R_E$. This can
        be converted to physical units when the lens star is detected.
\end{itemize}

For the remaining events with undetectable primary stars, the measurable
parameters are the following:

\begin{itemize}
  \item The planetary mass fraction, $\epsilon = M_{\rm planet}/M_\ast$,
        is determined from the microlensing light curve.
  \item The planet-star separation is 
        measured in units of the Einstein ring radius, $R_E$, and this
        can be converted to physical units with an accuracy of
        a factor of $\sim 2$.
  \item The masses of the free-floating planets must generally be
        determined from the event time scale only. This can be done
        to an accuracy of a factor of three for each individual event.
  \item Many of the $\sim 1\,\mearth$ planets and virtually all of the
        $\sim 0.1\,\mearth$ planets detected will have caustic crossing
        features which depend on the ratio of the source star radius to
        $R_E$. This will allow a mass estimate with an accuracy of a factor
        of two for planets orbiting a star or detected as isolated objects.
\end{itemize}

 
\subsection{Planet Detection via Transits}
\label{subsec-transit}

While the focus of the space mission that we propose is to find low mass 
planets via gravitational microlensing, the survey will also be sensitive to
giant planets via transits of the $\sim 10^8$ Galactic bulge
stars being monitored. Since giant planets like Jupiter
have a radius that is about 10\% of a solar radius, a
transit of a Jupiter-like planet across the Sun will
reduce the apparent brightness of the Sun by about 1\%.
The proposed GEST telescope
has the sensitivity to detect such a transit of a
solar-type Galactic bulge star by a Saturn size planet, and the
following argument shows that such a mission can
detect transits of Saturn size planets orbiting fainter
main sequence stars, as well. The luminosity and
radius of a main sequence star obeys the following
approximate relations: $ L\propto M^{3.5}$ and $R\propto M$.  Since the
fractional photometric signal from a transiting planet
(of a fixed radius) goes as $R^{-2}$, the signal-to-noise for
a transiting planet scales as $M^{-0.25}$, which is a very weak
dependence slightly favoring lower mass stars.

Some of the $\sim 10^8$ target stars will have images that are
blended with those of their near neighbor stars, and this can cause a 
substantial increase in the photon noise which significantly reduces the
sensitivity to planetary transits. This effect has been included in our
calculations of the expected numbers of detectable planetary transits.
The number of expected planetary transit detections for
planets at different orbital distances are
summarized in Table 1 which assumes a detection threshold of
a $6.5\sigma$ detection of a planet of Saturn's radius in 5 hours of
exposures. This translates into a $9\sigma$ detection of a Jupiter
sized planet. A crucial ingredient of our transit detection calculation
is the inclusion of
realistic stellar radii for the source stars, because many of them 
have a radius that is substantially smaller than the Sun.

Planets with orbital periods longer than 4 years can be detected via
transits, but only one transit will be detected per planet. Such
transits should have enough signal-to-noise for a significant detection
because the transit duration is $\simgt 10\,$hours,
but the period of the planet can only be roughly estimated from
the transit duration. Because of the huge number of stars that will be
observed, planets out to $\sim 20\,$AU are detectable even though there is
only a probability of $\sim 2\times 10^{-6}$ that such a planet would have
its orbit aligned with the line of sight and have
the right orbital phase to transit the source star during the period of
observations. This sensitivity to distant planets via transits means that
a space-based microlensing planet search mission will have a 
very substantial overlap in the planetary separations probed by the
microlensing and transit techniques. At orbital distances of 0.4-20$\,$AU,
the proposed GEST mission will be sensitive to giant
planets through both methods. This will allow cross-checks to 
help confirm the planetary interpretation of the transits. Since the
transit signal indicates radius rather than mass, some of the transits
could be caused by low mass M-dwarfs or brown dwarfs with similar radii,
but much larger masses than giant planets. Thus, some form
of confirmation is desirable. For example, we might measure the radial
velocities of some sub-sample of the candidate planets detected via
transits using a moderate resolution multi-object spectrograph. This 
would not allow us to distinguish between giant planets and low-mass
brown dwarfs, but we should detect radial velocity variations for 
for those stars which are transited by M-dwarfs or high-mass brown dwarfs.
This might allow a statistical correction for the non-planetary transits.

With the combined sample of microlensing and transit detections of giant
planets, a wide FOV space telescope
will be able to probe the entire range of giant planet
orbital radii: from 0, where the transit technique is very efficient, to
$\infty$, where microlensing is the only viable technique. Thus, such a
telescope promises a complete survey of giant planets with the 
combination of the two techniques.

\subsection{Additional Science with a wide FOV Space Telescope}
\label{subsec-add}

There are several other space-based microlensing planet search capabilities that we 
have not discussed in detail. Planets orbiting a single star of a binary
system have been detected via radial velocities \cite{marcy-but}, and
gravitational microlensing evidence has been presented for a planet
orbiting a binary star system \cite{mps-97blg41}, although this interpretation
remains uncertain \cite{planet-97blg41} due to incomplete coverage of
the microlensing light curve. 

An additional capability that we have not discussed in this paper is the
possibility of studying the abundance of planets in external galaxies,
such as M31 \cite{extra-gal-planet}. While most of the source stars in
M31 will be either poorly resolved or unresolved by a telescope with
angular resolution that is no better than that of HST, it is still possible to
detect microlensing events with giant star sources if the microlensing
magnification is not too small. Because an M31 planet search follows mostly
giant source stars, it will not be very sensitive to terrestrial extra-solar
planets, but it should be able to a detect large number of giant planets
at a separation of 1-10$\,$AU and measure their abundance as a function 
of position in the galaxy. 

Other possible science programs include a high redshift supernovae
search and a deep, wide-field, high resolution weak lensing survey.
(Both of these are goals of the proposed SNAP mission.) It would also
be possible to carry out a deep Kuiper Belt Object (KBO) search which 
should discover 100,000 new KBOs \cite{dps2000-kbo}. Many of these programs
could be carried out during the 4 months per year when the Galactic bulge
planet search field is too close to the Sun to be observed, and they
might be selected as a part of a general observer program
via a competitive review. 

\section{Conclusions}
\label{sec-conclude}

In this paper, we have presented the results of a simulations of 
a space-based gravitational microlensing survey for terrestrial 
extra-solar planets, similar to the
proposed GEST mission. We have determined the expected planet
detection sensitivity as a function of the planetary mass fraction, $\epsilon$,
and the orbital semi-major axis, and we have shown how the sensitivity
to Earth-like planets depends on the telescope parameters. 
We have found that such a mission will be sensitive
to planets down to a tenth of an Earth mass, or about 1000 times less than the
masses of planets discovered with the radial velocity technique. 

We have shown that a space-based microlensing planet search program should
be able to directly detect the planetary host (and lens)
stars for about one third of the
detectable planets. The observations of the host star when combined with the
microlensing light curve will allow the determination of the planetary mass
and separation as well as the stellar mass, type, and Galactocentric distance.
The visible stars include virtually all of the ``solar type" lens stars,
\ie\ those of spectral type F, G, or K which comprise about 25\% of the
total. For the remainder of the lens stars, which are mostly M-dwarfs,
it is generally
possible to accurately determine the planetary mass fraction and 
to determine the projected planet-star separation to an accuracy of a
factor of 2. 
For about one third of the detected planets, the lens star should
be directly detectable in the space-based survey data and with
ground-based infrared observations (with adaptive optics). This allows
an accurate determination of the mass and distance to the primary as
well as the planetary separation in physical units.

The expected scientific output of a space-based microlensing
planet search program is summarized here:
\begin{itemize}
  \item The average number of planets per star down to $0.1\,\mearth$
        at separations of $\sim0.7\,$AU - $\infty$ for terrestrial planets
        and 0 - $\infty$ for giant planets.
  \item The planetary mass function as a function of the planetary
        mass fraction, $f(M_{\rm planet}/M_\ast)$, and separation, for
        all lens stars.
  \item The planetary mass function as a function of stellar mass, 
        Galactocentric distance, and
        the planet-star separation for G, K, and early M stars.
  \item The abundance of giant planet pairs. A high abundance will indicate
        a large fraction of near circular orbits.
  \item The ratio of free-floating to bound planets as a function of planetary
        mass.
\end{itemize}

Finally, we would like to emphasize that the results that we have presented are
based upon very conservative assumptions. We've assumed a microlensing
optical depth number that is 1.3 times smaller than the latest measurements
indicate. If we assume that the optical depth measurement errors have
a normal distribution, this is the 95\% confidence level lower limit
on the microlensing optical depth. 

We've also been conservative in the selection of our planet selection
criteria by demanding a $12.5\sigma$ improvement ($\Delta\chi^2 \geq 160$)
for a planetary microlensing fit compared to a single lens fit. This 
ensures that we can make a reasonably accurate determination of the 
planetary parameters, but the event count could probably by increased
by about 70\% if the threshold was dropped to $9\sigma$. Furthermore,
events with a peak magnification $A_{\rm max} > 200$ have not been
included because they may be difficult to interpret. All told, if we dropped 
all of our conservative assumptions, we would have an event rate that
is 2-3 times higher than we have reported (although the interpretation of
some of these events might be difficult).

In summary, we've demonstrated that the a space-based microlensing planet
search mission can detect planets
with masses down to that of Mars which is a tenth of and Earth mass and
some three orders of magnitude better than current techniques. 
Space-based microlensing is unique among indirect terrestrial
planet search programs in that low mass planets are detected at high
signal-to-noise. Such a mission would be
sensitive to terrestrial planets
at orbital distances of $\simgt 0.7\,$AU via microlensing as well as
giant planets are all orbital radii via both microlensing and transits.
If each star has a $1\,\mearth$ planet orbiting at $1\,$AU, GEST would detect
$\sim 100$ of these. For about one third of the detected planets, the
host stars would be directly observable in the images. This will
allow the determination of the stellar type, mass, and distance, and it
will allow an accurate estimate of the planet-star separation in AU.
The results we've presented indicate that a space-based microlensing planet
search program
could provide very useful statistics on the abundance of terrestrial and
giant planets well in advance of the Terrestrial Planet Finder (TPF) mission,
and this information would likely be quite useful in planning TPF.

\acknowledgements
\section*{Acknowledgments}
We'd like to thank John Mather for encouragement and advice during the 
early stages of this work, and we'd also 
like to thank Domenick Tenerelli and his team at Lockheed Martin Space
Systems for their efforts on behalf of the GEST Discovery proposal.
Finally, we'd like to thank the GEST Co-Investigators for their help.
The Co-Investigators are: I. Bond, E. Cheng, J. Connor, K. Cook, 
P. Garnavich, K. Griest, D. Jewitt, N. Kaiser, T. Lauer, J. Lunine,
G. Luppino, D. Minniti, S. Peale, M. Shao, R. Stevenson, C. Stubbs,
N. Woolf, and P. Yock.
This work was supported, in part, by the NASA Origins
Grants NAG5-4573 and NAG5-9731.


\bibliographystyle{apj}

\onecolumn

\clearpage

\clearpage

\begin{figure}
\plottwo{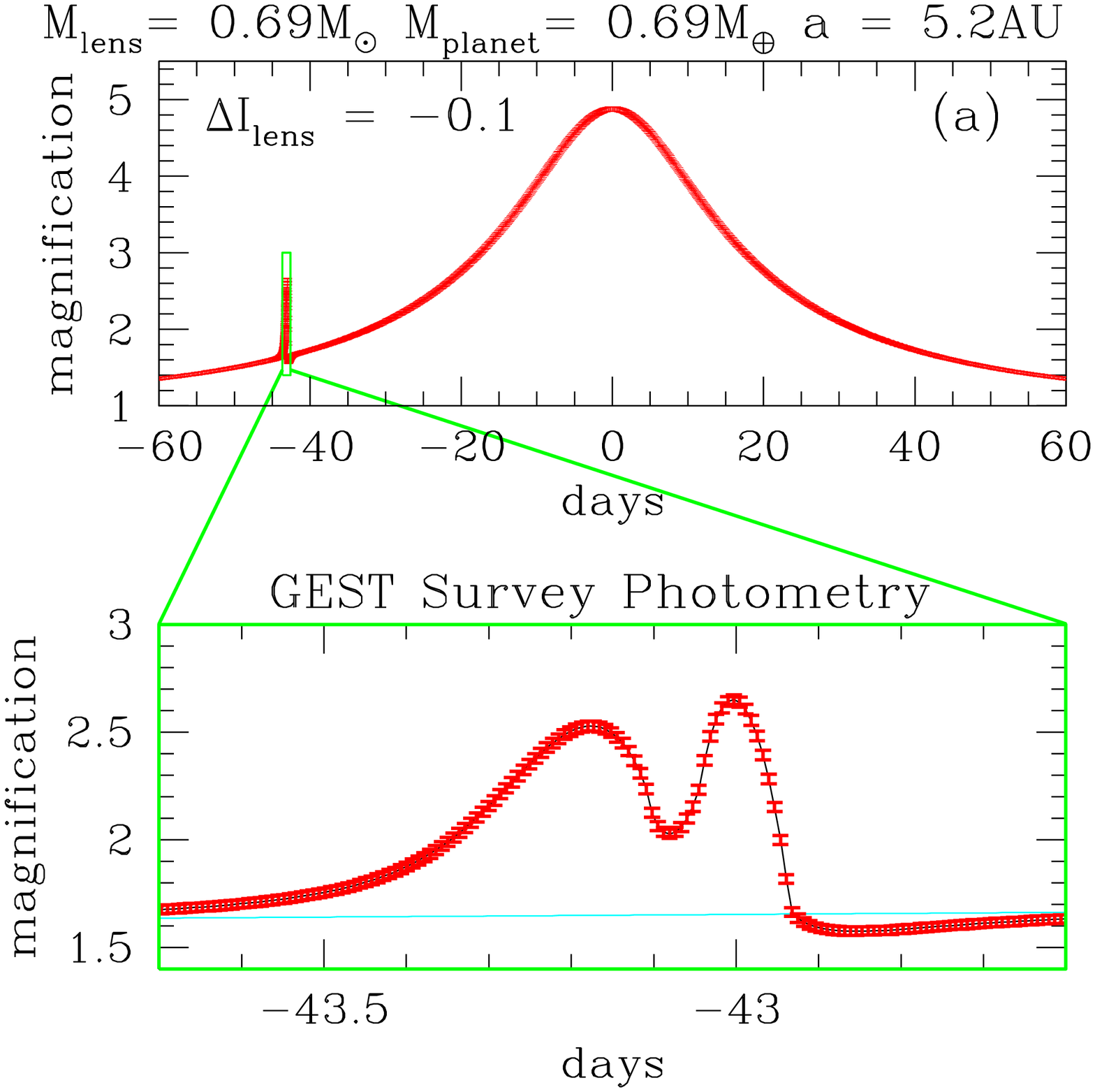}{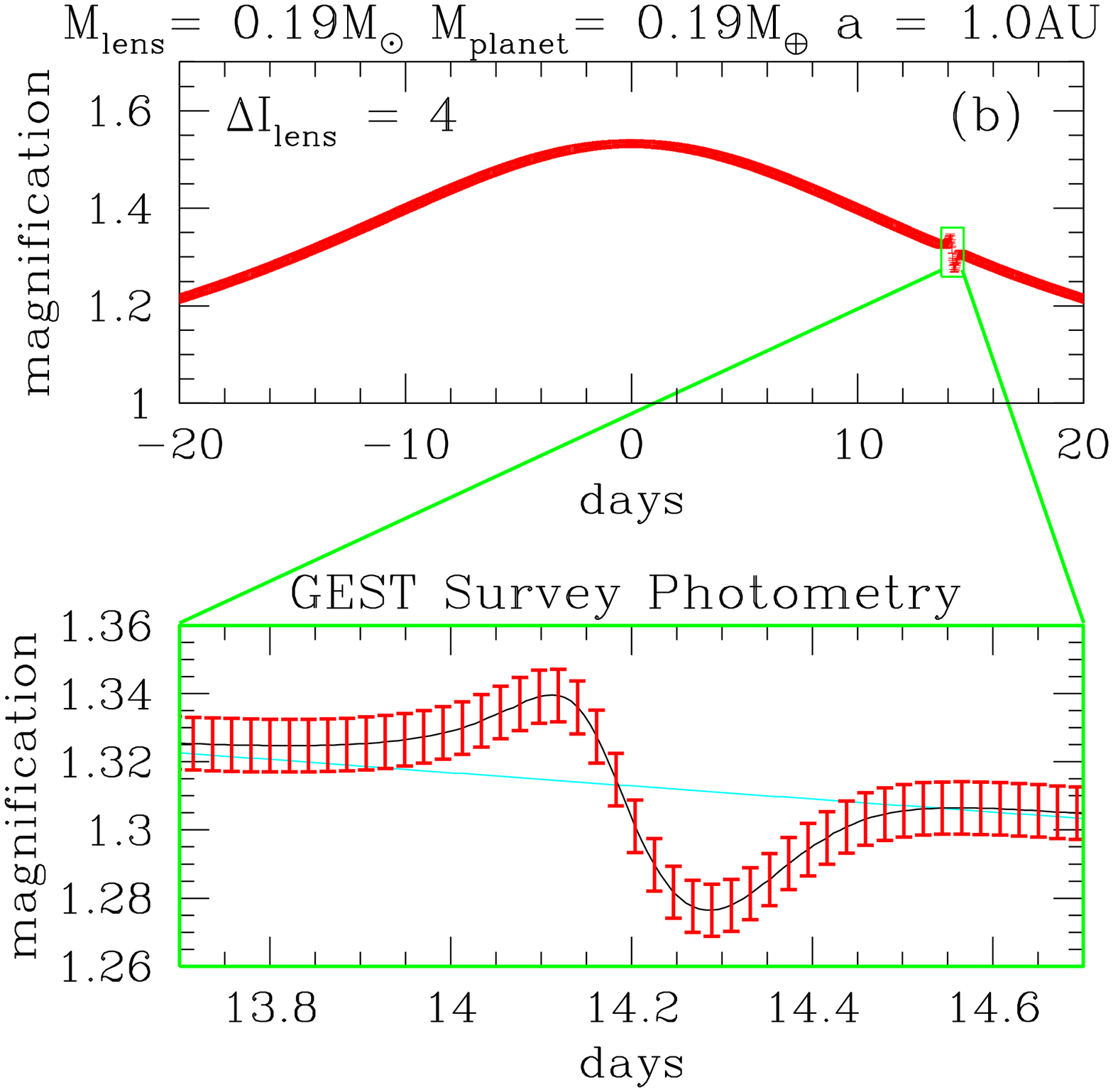}
\end{figure}

\clearpage

\clearpage

\clearpage
\begin{deluxetable}{cccccccc}
\tablecaption{Terrestrial Planet Detection Sensitivity for 
    $\epsilon = 3\times 10^{-6}$ at 0.7-1.5 AU}
\tablewidth{0pt}
\scriptsize
\tablehead{
  \colhead{$\epsilon$} &
  \colhead{FOV} &
  \colhead{minimum} &
  \colhead{FWHM} &
  \multicolumn{4}{c}{Effective Telescope Aperture (m)} \nl
  \colhead{} &
  \colhead{(sq. deg.)} &
  \colhead{error} &
  \colhead{} &
  \colhead{1.0} &      
  \colhead{\bf 1.25} &  
  \colhead{1.6} &      
  \colhead{2.0}        
}
\startdata
$3\times 10^{-6}$ & 1.5 & 0.3\% & 0.32" & 0.612 & 0.759 & 0.921 & 1.049 \nl
$3\times 10^{-6}$ & 2.0 & 0.3\% & 0.32" & 0.726 & 0.908 & 1.058 & 1.251 \nl
$3\times 10^{-6}$ & 2.5 & 0.3\% & 0.32" & 0.839 & 1.058 & 1.281 & 1.454 \nl
$3\times 10^{-6}$ & 1.5 & 0.15\% & 0.32" & 0.652 & 0.813 & 1.006 & 1.167 \nl
$3\times 10^{-6}$ & 2.0 & 0.15\% & 0.32" & 0.778 & 0.976 & 1.205 & 1.396 \nl
$3\times 10^{-6}$ & 2.5 & 0.15\% & 0.32" & 0.905 & 1.138 & 1.405 & 1.625 \nl
$3\times 10^{-6}$ & 1.5 & 0.3\% & 0.24" & 0.665 & 0.837 & 0.983 & 1.119 \nl
$3\times 10^{-6}$ & {\bf 2.0} & {\bf 0.3\%} & {\bf 0.24"} & 0.795 & {\bf 1.000} & 1.179 & 1.336 \nl
$3\times 10^{-6}$ & 2.5 & 0.3\% & 0.24" & 0.925 & 1.163 & 1.374 & 1.553 \nl
$3\times 10^{-6}$ & 1.5 & 0.15\% & 0.24" & 0.704 & 0.889 & 1.071 & 1.251 \nl
$3\times 10^{-6}$ & 2.0 & 0.15\% & 0.24" & 0.846 & 1.065 & 1.285 & 1.494 \nl
$3\times 10^{-6}$ & 2.5 & 0.15\% & 0.24" & 0.988 & 1.241 & 1.499 & 1.737 \nl
$3\times 10^{-6}$ & 1.5 & 0.3\% & 0.16" & 0.723 & 0.904 & 1.072 & 1.217 \nl
$3\times 10^{-6}$ & 2.0 & 0.3\% & 0.16" & 0.865 & 1.079 & 1.278 & 1.448 \nl
$3\times 10^{-6}$ & 2.5 & 0.3\% & 0.16" & 1.008 & 1.254 & 1.485 & 1.679 \cr
$3\times 10^{-6}$ & 1.5 & 0.15\% & 0.16" & 0.763 & 0.956 & 1.152 & 1.331 \nl
$3\times 10^{-6}$ & 2.0 & 0.15\% & 0.16" & 0.918 & 1.144 & 1.378 & 1.587 \nl
$3\times 10^{-6}$ & 2.5 & 0.15\% & 0.16" & 1.073 & 1.332 & 1.604 & 1.844 \cr
\enddata
\tablenotetext{} { 
This table shows the ratio of the number of terrestrial planet detections
as a function of the telescope aperture, field-of-view (FOV), and 
effective point spread function FWHM. The parameters of the GEST 
MIDEX proposal are indicated in {\bf bold}.
}
\end{deluxetable}
\clearpage
\begin{deluxetable}{cccccccc}
\tablecaption{Terrestrial Planet Detection Sensitivity for 
      $epsilon = 10^{-5}$ at 0.7-1.5 AU}
\tablewidth{0pt}
\scriptsize
\tablehead{
  \colhead{$\epsilon$} &
  \colhead{FOV} &
  \colhead{minimum} &
  \colhead{FWHM} &
  \multicolumn{4}{c}{Effective Telescope Aperture (m)} \nl
  \colhead{} &
  \colhead{(sq. deg.)} &
  \colhead{error} &
  \colhead{} &
  \colhead{1.0} &      
  \colhead{\bf 1.25} &  
  \colhead{1.6} &      
  \colhead{2.0}        
}
\startdata
$        10^{-5}$ & 1.5 & 0.3\% & 0.32" & 0.680 & 0.793 & 0.906 & 1.016 \nl
$        10^{-5}$ & 2.0 & 0.3\% & 0.32" & 0.803 & 0.937 & 1.074 & 1.202 \nl
$        10^{-5}$ & 2.5 & 0.3\% & 0.32" & 0.925 & 1.082 & 1.243 & 1.389 \nl
$        10^{-5}$ & 1.5 & 0.15\% & 0.32" & 0.709 & 0.834 & 0.974 & 1.114 \nl
$        10^{-5}$ & 2.0 & 0.15\% & 0.32" & 0.837 & 0.985 & 1.153 & 1.320 \nl
$        10^{-5}$ & 2.5 & 0.15\% & 0.32" & 0.965 & 1.136 & 1.322 & 1.526 \nl
$        10^{-5}$ & 1.5 & 0.3\% & 0.24" & 0.732 & 0.846 & 0.969 & 1.095 \nl
$        10^{-5}$ & {\bf 2.0} & {\bf 0.3\%} & {\bf 0.24"} & 0.863 & {\bf 1.000} & 1.148 & 1.296 \nl
$        10^{-5}$ & 2.5 & 0.3\% & 0.24" & 0.994 & 1.154 & 1.326 & 1.497 \nl
$        10^{-6}$ & 1.5 & 0.15\% & 0.24" & 0.758 & 0.885 & 1.034 & 1.190 \nl
$        10^{-5}$ & 2.0 & 0.15\% & 0.24" & 0.894 & 1.046 & 1.225 & 1.411 \nl
$        10^{-5}$ & 2.5 & 0.15\% & 0.24" & 1.031 & 1.207 & 1.415 & 1.632 \nl
$        10^{-5}$ & 1.5 & 0.3\% & 0.16" & 0.769 & 0.889 & 1.039 & 1.164 \nl
$        10^{-5}$ & 2.0 & 0.3\% & 0.16" & 0.906 & 1.049 & 1.230 & 1.372 \nl
$        10^{-5}$ & 2.5 & 0.3\% & 0.16" & 1.042 & 1.209 & 1.421 & 1.582 \cr
$        10^{-5}$ & 1.5 & 0.15\% & 0.16" & 0.794 & 0.928 & 1.101 & 1.256 \nl
$        10^{-5}$ & 2.0 & 0.15\% & 0.16" & 0.936 & 1.094 & 1.302 & 1.483 \nl
$        10^{-5}$ & 2.5 & 0.15\% & 0.16" & 1.078 & 1.261 & 1.504 & 1.712 \cr
\enddata
\tablenotetext{} { 
This table shows the ratio of the number of terrestrial planet detections
as a function of the telescope aperture, field-of-view (FOV) and 
effective point spread function FWHM. The parameters of the GEST 
MIDEX proposal are indicated in {\bf bold}.
}
\end{deluxetable}

\clearpage
\begin{deluxetable}{ccrrc}
\tablecaption{Planetary Transits from GEST}
\tablewidth{0pt}
\scriptsize
\tablehead{
  \colhead{Semi-major axis (AU)} &
  \colhead{Period (yrs.)} &
  \colhead{\# of detections} &
  \colhead{transits per planet} &
  \colhead{transit duration}
}
\startdata
0.04 &  $\sim 0.01$  &  5,000,000  &  $\sim 200$ &  1.6  \nl
0.4  &  $\sim 0.3$   &    600,000  &    $\sim 7$ &  5    \nl
1.0  &  $\sim 1.3$   &    160,000  &    $\sim 2$ &  8    \nl
2.0  &  $\sim 3.7$   &     40,000  &          1  & 11    \nl
5.2  &  $\sim 15$    &      6,000  &          1  & 18    \nl
9.5  &  $\sim 40$    &      1,300  &          1  & 24    \nl
19.5 &  $\sim 110$   &        200  &          1  & 35    \cr
\enddata
\tablenotetext{} { 
This table shows the number of expected transit planet detections for 
planets with a radius at least as large as that of Saturn for a three
year GEST mission assuming 8 months of observations per year. The planet
detection numbers assume 1 planet per star. 
}
\end{deluxetable}

\end{document}